\documentclass[structabstract]{aa}
\usepackage{graphicx}
\usepackage{color,soul}
\usepackage[varg]{txfonts}
\usepackage{longtable}
\usepackage{supertabular}
\usepackage{natbib}
\begin{document}

\newcommand{\ABi}[1]{\textcolor{magenta}{\bf Andrea says: #1}}
\newcommand{\MUl}[1]{\textcolor{red}{\bf Mel says: #1}}
\newcommand{\gam}[1]{\textcolor{blue}{\bf Gary: #1}}
\newcommand{\cvir}{c_{200}}
\newcommand{\msun}{M_{\odot}}
\newcommand{\mvir}{M_{200}}
\newcommand{\rvir}{r_{200}}
\newcommand{\rtr}{r_n}
\newcommand{\rs}{r_{-2}}
\newcommand{\slos}{\sigma_{\rm{los}}}

\def\trt{t_{\rm RT}}
\def\mathnew{\mathsurround=0pt}
\def\simov#1#2{\lower .5pt\vbox{\baselineskip0pt \lineskip-.5pt
\ialign{$\mathnew#1\hfil##\hfil$\crcr#2\crcr\sim\crcr}}}
\def\simgreat{\mathrel{\mathpalette\simov >}}
\def\simless{\mathrel{\mathpalette\simov <}}
\def\MeV{Me\kern-0.11em V}
\def\keV{ke\kern-0.11em V}
\def\arcsec{\hbox{$^{\prime\prime}$}}
\def\etal{{\it et~al.\/}}
\def\cha{{\it Chandra\/}}
\def\xmm{{\it XMM Newton\/}}
\def\pspc{{\it ROSAT/PSPC \/}}
\def\xl{{\it XMM LSS\/}}
\def\hst{{\it HST\/}}
\def\spi{{\it Spitzer\/}}
\def\z{{\it z\/}}
\def\Ms{M$_{\bigodot}$\/}

\title{The mass profile and dynamical status of the $z \sim 0.8$
  galaxy cluster LCDCS~0504~\thanks{Based on XMM-Newton archive data
    and on data retrieved from the NASA/IPAC Extragalactic Database
    (NED) which is operated by the Jet Propulsion Laboratory,
    California Institute of Technology, under contract with the
    National Aeronautics and Space Administration. Also based on
    observations made with the FORS2 multi-object spectrograph mounted
    on the Antu VLT telescope at ESO-Paranal Observatory (programme
    175.A-0706(B)). Also based on observations obtained at the Gemini
    Observatory, which is operated by the Association of Universities
    for Research in Astronomy, Inc., under a cooperative agreement
    with the NSF on behalf of the Gemini partnership: the National
    Science Foundation (United States), the Science and Technology
    Facilities Council (United Kingdom), the National Research Council
    (Canada), CONICYT (Chile), the Australian Research Council
    (Australia), Minist\'erio da Ci\^encia, Tecnologia e Inova\c c\~ao
    (Brazil) and Ministerio de Ciencia, Tecnolog\'\i a e Inovaci\'on
    Productiva (Argentina). Finally, this research has made use of the
    VizieR catalog access tool, CDS, Strasbourg, France.  Also based
    on observations made with the NASA/ESA Hubble Space Telescope,
    obtained from the data archive at the Space Telescope Science
    Institute. STScI is operated by the Association of Universities
    for Research in Astronomy, Inc. under NASA contract NAS 5-26555.
    Also based on visiting astronomer observations, at Cerro Tololo
    Inter-American Observatory, National Optical Astronomy
    Observatory, which is operated by the Association of Universities
    for Research in Astronomy, under contract with the National
    Science Foundation. This work has been carried out thanks to the
    support of the Labex OCEVU (ANR-11-LABX-0060) and the A*MIDEX
    (ANR-11-IDEX-0001-02) funded by the "Investments for the Future"
    French government program managed by the French National Research
    Agency (ANR)}}

\author{L.~Guennou\inst{\ref{LGu},\ref{CAd}} \and
A.~Biviano\inst{\ref{ABi},\ref{GMa}} \and
C.~Adami\inst{\ref{CAd}} \and
M. Limousin\inst{\ref{CAd}} \and
G.B.~Lima Neto\inst{\ref{ECy}} \and
G.A.~Mamon\inst{\ref{GMa}} \and
M.P.~Ulmer\inst{\ref{MUl}} \and
R.~Gavazzi\inst{\ref{GMa}} \and
E.S. Cypriano\inst{\ref{ECy}} \and
F.~Durret\inst{\ref{GMa}} \and
D.~Clowe\inst{\ref{DCl}} \and
V.~LeBrun\inst{\ref{CAd}} \and
S.~Allam\inst{\ref{SAl},\ref{SAl2}} \and
S.~Basa\inst{\ref{CAd}} \and
C.~Benoist\inst{\ref{CBe}} \and
A.~Cappi\inst{\ref{ACa},\ref{CBe}} \and
C.~Halliday\inst{\ref{CHa}} \and
O.~Ilbert\inst{\ref{CAd}} \and
D.~Johnston\inst{\ref{SAl}} \and
E.~Jullo\inst{\ref{CAd}} \and
D.~Just\inst{\ref{DJu},\ref{DJu2}} \and
J.M.~Kubo\inst{\ref{SAl}} \and
I.~M\'arquez\inst{\ref{IMa}}\and
P.~Marshall\inst{\ref{PMa}} \and
N.~Martinet\inst{\ref{GMa}} \and
S.~Maurogordato\inst{\ref{CBe}} \and
A.~Mazure\inst{\ref{CAd}} \and
K.J.~Murphy\inst{\ref{DCl}} \and
H. Plana\inst{\ref{HPl}} \and
F.~Rostagni\inst{\ref{CBe}} \and
D.~Russeil\inst{\ref{CAd}} \and
M.~Schirmer \inst{\ref{MSc},\ref{MSc2}} \and 
T.~Schrabback\inst{\ref{TSc}} \and
E.~Slezak\inst{\ref{CBe}} \and
D.~Tucker\inst{\ref{SAl}} \and
D.~Zaritsky\inst{\ref{DJu}} \and
B.~Ziegler \inst{\ref{BZi}}
}

\offprints{L.~Guennou, guennou@ukzn.ac.za}

\institute{Astrophysics and Cosmology Research Unit, 
University of KwaZulu-Natal, Durban, 4041, South Africa 
\label{LGu} \and
Aix Marseille Universit\'e, CNRS, LAM (Laboratoire d'Astrophysique 
de Marseille) UMR 7326, 13388, Marseille, France 
\label{CAd} \and
INAF/Osservatorio Astronomico di Trieste, via Tiepolo 11, 
34143 Trieste, Italy 
\label{ABi} \and
Institut d'Astrophysique de Paris (UMR7095: CNRS \& UPMC), 98 bis Bd Arago, F-75014, 
Paris, France 
\label{GMa} \and
Departamento de Astronomia, Instituto de Astronomia Geof\`isica e Ci\^encias 
Atmosf\`ericas, Universidade de S\~ao Paulo, Rua do Mat\~ao 1226, 05508-900 S\~ao 
Paulo, Brazil 
\label{ECy} \and
Department Physics \& Astronomy \& Center for Interdisciplinary Exploration
and Research in Astrophysics (CIERA), 
Northwestern University, Evanston, 
IL 60208-2900, USA 
\label{MUl} \and
Department of Physics and Astronomy, Ohio University, 251B Clippinger
Lab, Athens, OH 45701, USA 
\label{DCl} \and
Fermi National Accelerator Laboratory, P.O. Box 500, Batavia, IL 60510, USA 
\label{SAl} \and
CSC/STSCi, 3700 San Martin Dr., Baltimore, MD 21218, USA
\label{SAl2} \and
OCA, Cassiop\'ee, Boulevard de l'Observatoire, BP 4229, 06304 Nice Cedex 4, France 
\label{CBe} \and
INAF/Osservatorio Astronomico di Bologna, via Ranzani 1, 40127 Bologna, Italy
\label{ACa} \and
23, rue d'Yerres, 91230 Montgeron, France
\label{CHa} \and
Steward Observatory, University of Arizona, 933 N. Cherry Ave. Tucson, 
AZ 85721, USA
\label{DJu} \and
Department of Astronomy \& Astrophysics, University of Toronto, 50 St
George Street Toronto, Ontario M5S 3H4, Canada
\label{DJu2} \and
Instituto de Astrof\'\i sica de Andaluc\'\i a (CSIC), Glorieta de la 
Astronom\'\i a s/n, 18008, Granada, Spain 
\label{IMa} \and
Kavli Institute for Particle Astrophysics and Cosmology, Stanford University, 
2575 Sand Hill Rd., Menlo Park, CA 94025, USA
\label{PMa} \and
Laborat\'orio de Astrof\'\i sica Te\'orica e Observacional, Universidade
Estadual de Santa Cruz, Ilh\'eus, Brazil
\label{HPl} \and
Gemini Observatory, Casilla 603, La Serena, Chile
\label{MSc} \and
Argelander-Institut f\"ur Astronomie, Universit\"et Bonn, Auf dem
H\"ugel 71, 53121, Bonn, Germany
\label{MSc2} \and
Physics Department, University of California, Santa Barbara, CA 93601, USA 
\label{TSc} \and
University of Vienna, Department of Astrophysics, T\"urkenschanzstr. 17, 1180 Wien, Austria
\label{BZi}
}

\date{Accepted . Received ; Draft printed: \today}

\authorrunning{L. Guennou et al}

\titlerunning{Dynamics of the cluster LCDCS~0504}

\abstract{Constraints on the mass distribution in high-redshift
  clusters of galaxies are not currently very strong.}{We aim to
  constrain the mass profile, $M(r)$, and dynamical status of the $z
  \sim 0.8$ LCDCS~0504 cluster of galaxies characterized by prominent
  giant gravitational arcs near its center.}  {Our analysis is based
  on deep X-ray, optical, and infrared imaging, as well as optical
  spectroscopy, collected with various instruments, complemented with
  archival data. We model the mass distribution of the cluster with
  three different mass density profiles, whose parameters are
  constrained by the strong lensing features of the inner cluster
  region, by the X-ray emission from the intra-cluster medium, and by
  the kinematics of 71 cluster members.}  {We obtain consistent $M(r)$
  determinations from three methods based on kinematics
  (dispersion-kurtosis, caustics and MAMPOSSt), out to the cluster
  virial radius, $\simeq 1.3$ Mpc and beyond. The mass profile
  inferred by the strong lensing analysis in the central cluster
  region is slightly above, but still consistent with, the
  kinematics estimate. On the other hand, the X-ray based $M(r)$ is
  significantly below both the kinematics and strong lensing
    estimates. Theoretical predictions from $\Lambda$CDM cosmology
  for the concentration--mass relation are in agreement with our
  observational results, when taking into account the uncertainties in
  both the observational and theoretical estimates.  There appears to
  be a central deficit in the intra-cluster gas mass fraction compared
  to nearby clusters.}  {Despite the relaxed appearance of this
  cluster, the determinations of its mass profile by different probes
  show substantial discrepancies, the origin of which remains to be
  determined. The extension of a similar dynamical analysis to other
  clusters of the DAFT/FADA survey with multi-wavelength data of
  sufficient quality, will allow to shed light on the possible
  systematics that affect the determination of mass profiles of
  high-$z$ clusters, possibly related to our incomplete understanding
  of intracluster baryon physics.}


\maketitle

\section{Introduction}
\label{s:intro}
The study and characterization of the internal dynamics of galaxy
clusters is an important way to understand their evolutionary history,
which is itself related to the evolutionary history of the
universe. The most classical way to characterize the dynamics of
clusters is through the analysis of the projected phase space
distribution of their member galaxies, e.g. via methods based on the
Jeans equation \citep{BT87}, such as the Dispersion-Kurtosis
\citep{LM03}, distribution-function \citep{Wojtak+09} and MAMPOSSt
\citep{MBB13} methods, or the Caustic method calibrated on numerical
simulations \citep{DG97}. All these methods assume spherical symmetry
and most of them (except the Caustic method) also assume dynamical
relaxation of the cluster. These methods have been applied to several
nearby (and massive) clusters of galaxies
\citep[see][]{KG82,vanderMarel+00,BG03,LM03,BK04,KBM04,Biviano06b,Lokas+06,WL10}.

\begin{figure*}
\begin{center}
\includegraphics[width=13cm]{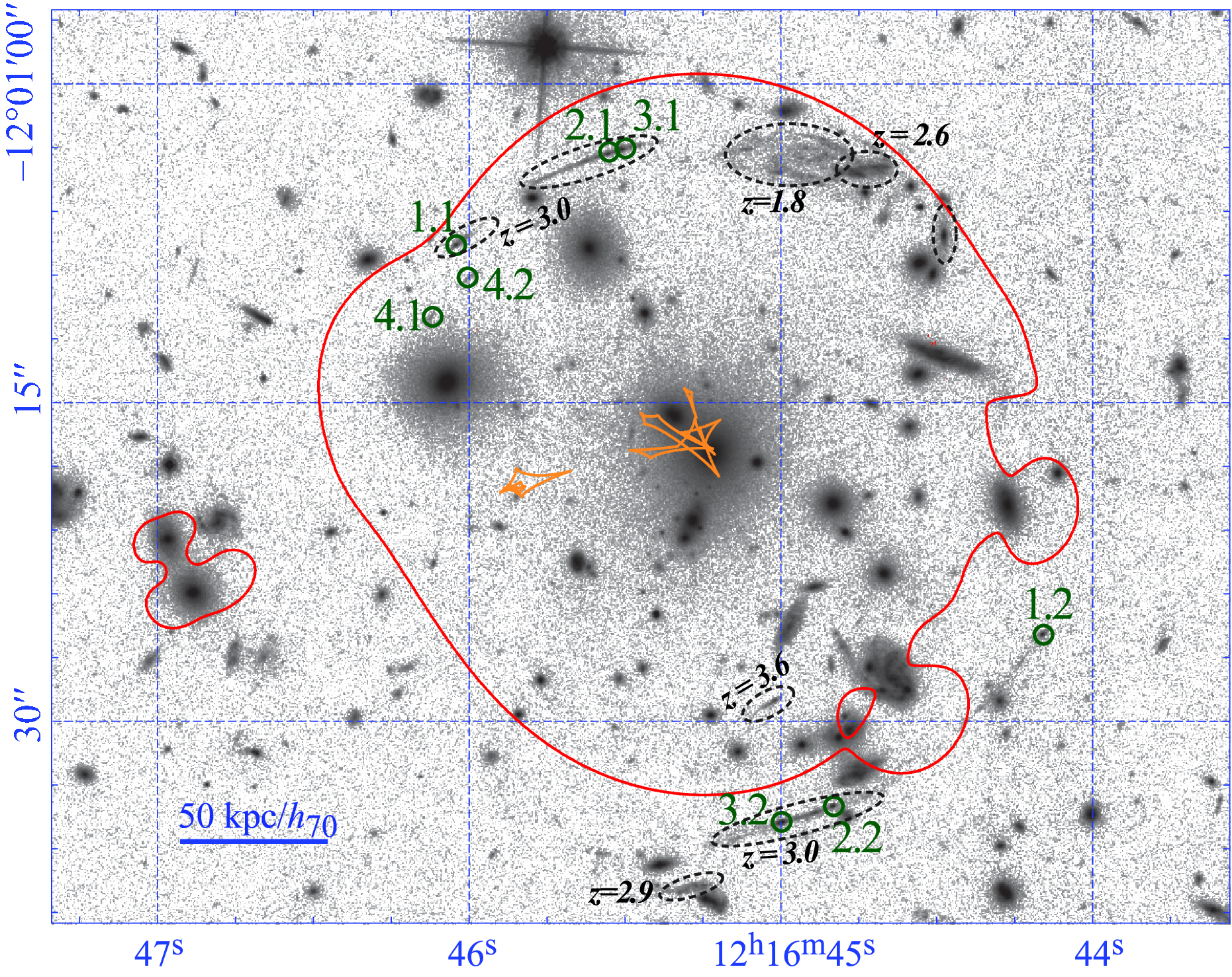}
\end{center}
\caption{HST image of the core of LCDCS~0504. The size of the field is 38$\times$34
  arcsec$^{2}$, corresponding to $285 \times 255$~kpc$^{2}$ at $z = 0.794$.  Multiple
  imaged systems used in this work are labeled. From the best fit strong lensing model, we draw in red the tangential critical curve at z=3 and the corresponding caustic lines in orange}
\label{f:multiple}
\end{figure*}

Given that clusters formed relatively recently according to the
hierarchical scenario of structure evolution in the universe
\citep[e.g.][]{BG01}, accretion of matter from the surrounding field,
in the form of galaxy groups, complicate their internal
structure. Detection of secondary structures, or substructures, in
clusters is obtained using other methods, either based on the
projected distributions of cluster galaxies
\citep[e.g.][]{DS88,Escalera+94,Biviano+96,SG96,Barrena+02,GB02,Ramella+07}
or on X-ray data for the intra-cluster gas
\citep{Briel+91,MFG93,Neumann+01,OHara+04,PBF05,Bohringer+10}. Detection
and characterization of these substructures is a direct way to
constrain the cluster building history \citep[e.g.][ and references
  therein]{Adami+05}.

These last years, the characterization of the mass distribution and
substructures of galaxy clusters has been made possible by
investigating deep and high quality data that enable the
  measurement of weak lensing signal and the detection of strong
  gravitationally lensed features
\citep[e.g.][]{Cypriano+04,Markevitch+04,Bardeau+05,Jee+05,Coe+10,LKG11}. It
is still relatively uncommon to see cluster dynamical studies based
simultaneously on the Jeans analysis, and on the X-ray and lensing
data, especially for high redshift clusters. This is due to the
extreme difficulty in obtaining both deep and high resolution X-ray
imaging, deep optical and infrared imaging, and faint galaxy
spectroscopy. As a consequence, our information on the internal
structure and dynamics of distant clusters is still relatively
limited.

In this paper, we perform a detailed study of the internal structure
and dynamics of the rich cluster LCDCS~0504 at redshft $z=0.7943$,
also known as Cl~J1216.8-1201 \citep{Nelson+01}, using simultaneously
spectroscopic optical data for cluster galaxies, as well as X-ray and
strong lensing (SL, hereafter) data. This cluster is part of the
DAFT/FADA survey \citep{Guennou+10} and the analysis presented here is
a proof of concept for similar analysis to be performed on other
clusters of the DAFT/FADA sample.

In Sect.~\ref{s:data} we present our data-set. Our SL determination of
the cluster mass distribution is described in Sect.~\ref{s:sl}. In
Sect.~\ref{s:x} we use the X-ray emission from the hot intra-cluster
medium (ICM) to constrain the cluster mass profile. This is also
determined using galaxies as tracers in Sect.~\ref{s:kin}.  We compare
the different mass profile determinations in Sect.~\ref{s:cmp}.  In
Sect.~\ref{s:bary} we analyse the cluster hot gas mass fraction.  We
discuss our results in Sect.~\ref{s:summ}, where we also draw our
conclusions.

Throughout this paper we adopt $H_0=70$ km~s$^{-1}$~Mpc$^{-1}$,
$\Omega_m=0.3$, $\Omega_{\Lambda}=0.7$. In this cosmology, 1 arcmin
corresponds to 449 kpc at the cluster redshift.

\begin{table}
\begin{center}
\caption{Available data for the LCDCS~0504 cluster.}
\label{t:datasumm}
\begin{tabular}{lcc}
\hline
& Archival data & DAFT/FADA data \\
\hline
& & \\
Optical imaging & $VRIz$ (VLT/FORS2) & B (Blanco/MOSAIC) \\
                             & $F814W$ (HST/ACS)  & \\
& & \\
IR imaging & Spitzer/IRAC1 and 2 & \\
& & \\
Optical & VLT/FORS2 & Gemini/GMOS \\
spectroscopy & & \\
& & \\
 X-ray imaging & XMM-Newton & \\
& (PN/MOS1/MOS2) & \\
\hline
\smallskip
\end{tabular}
\end{center}
\end{table}

\section{The data}
\label{s:data}

The DAFT/FADA survey is described at http://cesam.oamp.fr/DAFT/. Here
we focus on the description of the data available for LCDCS~0504,
summarized in Table~\ref{t:datasumm}.

\subsection{Optical and near-infrared imaging}
\label{ss:imaging}
We refer to \citet{Guennou+10} for a complete description of the
optical and infrared imaging data and  for the evaluation of photometric
redshifts, $z_{\rm p}$. These photometric redshifts are characterized by typical
uncertainties lower than 0.1 up to $z \sim$1.5 for galaxies
brighter than $F814W=22.5$, or up to $z \sim$1 for galaxies brighter than
$F814W=24$. The photometric redshifts are used here to define
cluster membership in the absence of spectroscopic information (see
Sect.~\ref{ss:nr}). 

\subsection{Optical spectroscopy}
\label{ss:spectro}
\begin{figure}
\begin{center}
\includegraphics[width=6truecm,angle=-90.0]{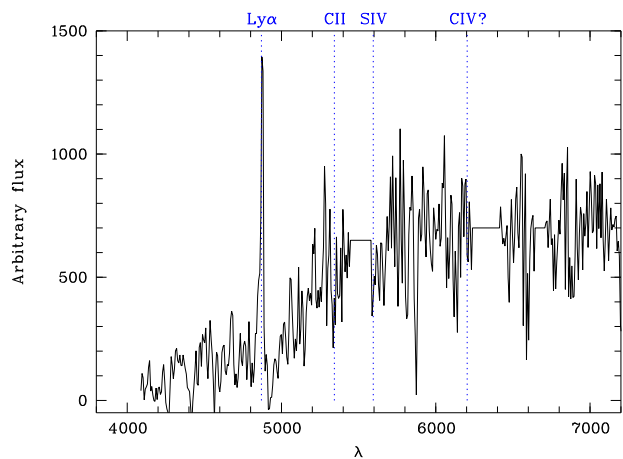}
\end{center}
\caption{Summed spectrum of arcs 1.1, 1.2, and 1.3. The best redshift is 3.005.}
\label{f:arc}
\end{figure}

We collected 116 galaxy redshifts from the NED database, originally
from \citet{Halliday+04}, obtained with VLT/FORS2 observations in a 5
arcmin radius around the cluster center. The average error on these
redshift measurements corresponds to 90 km~s$^{-1}$ in velocity. This
sample of spectroscopic redshifts is limited to $z \leq 1.1$. The
magnitude distribution of the spectroscopic sample peaks at an
$I$-band magnitude of 22, and is limited to $I \leq 24$.
\sethlcolor{yellow}
We were awarded 6 hours of Gemini/GMOS time (program GS2011A-014) to observe
spectroscopically the three brightest giant arcs.  The initial
spectral resolution was 150, but was degraded to 10~\AA/px in order to
increase the S/N. This theoretically provides a redshift uncertainty
of the order of 0.0015 (i.e. $\sim 500$~km~s$^{-1}$), corresponding
to an uncertainty of 1~pixel in the line location. Following the
identification labels of the observed objects in
Fig.~\ref{f:multiple}, we measured $z \sim 2.988$ for the object
1.3/1.4, $z \sim 3.005$ for the object 1.2, and $z \sim 3.009$ for the
object 1.1.  Assuming that the 3 objects are multiple images of a
single background object, we stacked the 3 spectra together (see
Fig.~\ref{f:arc}) and measured a redshift of 3.005 for the stacked
spectrum. From the best fit strong lensing model, we draw in red the tangential critical curve at z=3 (location where the amplification diverges) and in orange the caustic lines (which are generated by de-lensing the critical lines in the source plane)

\onltab{
\begin{table*}
\begin{center}
\caption{Coordinates, magnitudes, and redshifts of the
  LCDCS~0504 spectroscopic galaxy catalogue. Asterisks mark data from
  our GMOS run.}
\begin{tabular}{cccc|cccc}
\hline
RA (J2000) & Dec (J2000) & $I$ & redshift        & RA (J2000) & Dec (J2000) & $I$ & redshift \\
\hline
  12   16   35.84  &--12  03    16.4  & 22.20   & 0.7850    &    12   16   44.70  &--12  01    28.2  & 21.28 & 0.7865  \\
  12   16   35.90   &--12  00    29.4  & 21.62 & 0.7930 	 &    12   16   44.72  &--12  01    23.4  & 22.25 & 0.7945  \\
  12   16   36.13 &--12  00    43.8 & 22.04 & 0.6740*	 &    12   16   44.74  &--11  59   16.2  & 22.10 & 0.7998  \\
  12   16   36.14  &--11  59   01.4   & 21.52 & 0.4816 	 &    12   16   44.84  &--12  01    30.9  & 21.97 & 0.7984  \\
  12   16   36.27  &--12  03    29.0  & 21.34 & 0.5894 	 &    12   16   44.87  &--12  01    20.3  & 21.54 & 0.8035  \\
  12   16   36.37 &--11  59   20.0 & 22.74 & 0.6650*	 &    12   16   44.87  &--12  00    43.5  & 22.93 & 0.7824  \\
  12   16   36.41  &--12  00   08.7   & 21.64 & 0.7868 	 &    12   16   44.91  &--12  02    13.9  & 20.90 & 0.6691  \\
  12   16   36.51 &--12  00    31.9 & 22.38 & 0.6740*	 &    12   16   44.91  &--12  02    03.6   & 21.50 & 0.7938  \\
  12   16   36.54  &--12  02    29.8  & 22.90   & 0.4700*	 &    12   16   45.09  &--11  58   49.3  & 21.18 & 0.7969  \\
  12   16   36.62 &--12  02    29.8 & 22.00 & 0.4700*	 &    12   16   45.12  &--12  00    35.9  & 23.02 & 0.7883  \\
  12   16   37.18  &--12  00    41.9  & 21.03 & 0.6606 	 &    12   16   45.18  &--11  58   20.0  & 21.54 & 0.2327  \\
  12   16   37.74  &--12  03    48.6  & 21.72 & 0.7940 	 &    12   16   45.24  &--12  03    13.4  & 21.60 & 0.7933  \\
  12   16   38.01 &--12  02    51.4 & 21.89 & 0.7900*	 &    12   16   45.26  &--12  01    17.6  & 20.66 & 0.7955  \\
  12   16   38.12  &--12  03    26.6  & 21.26 & 0.7939 	 &    12   16   45.32  &--12  01    20.9  & 21.30 & 0.8054  \\
  12   16   38.23  &--12  02    51.7  & 21.08 & 0.7900 	 &    12   16   45.37  &--12  00    01.7   & 21.70 & 0.7996  \\
  12   16   38.40   &--11  59   15.2  & 20.62 & 0.2758 	 &    12   16   45.60  &--11  58   38.3  & 22.85 & 0.7925  \\
  12   16   38.74  &--12  01    50.3  & 21.50 & 0.8008 	 &    12   16   45.65  &--12  01    08.0   & 21.64  & 0.8058  \\
  12   16   38.74  &--12  03    12.0  & 21.19 & 0.7958 	 &    12   16   45.83  &--12  01    05.6   & 22.94 & 0.7921  \\
  12   16   38.83  &--12  02    44.2  & 20.79 & 0.4167 	 &    12   16   45.91  &--12  03    29.4  & 21.69 & 0.2252  \\
  12   16   39.08 &--12  00    15.6 & 22.40   & 0.8890*	 &    12   16   46.10  &--12  01    14.3  & 20.61 & 0.7997  \\
  12   16   39.08  &--12  03    35.7  & 22.55 & 0.6601 	 &    12   16   46.18  &--12  02    25.3  & 21.55 & 0.7866  \\
  12   16   39.11 &--11  58   53.6 & 21.59 & 0.6640*	 &    12   16   46.20  &--12  00    31.0  & 22.27 & 0.7952  \\
  12   16   39.13 &--11  58   39.9 & 21.60 & 0.6650*	 &    12   16   46.23  &--12  00    07.3   & 22.06 & 0.7847  \\
  12   16   39.24  &--11  58   03.4   & 22.41 & 0.8816 	 &    12   16   46.35  &--12  03    25.7  & 22.34 & 0.7966  \\ 
  12   16   39.41  &--12  03    46.4  & 21.05 & 0.5888 	 &    12   16   46.39  &--11  59   34.0  & 22.64 & 0.7936  \\ 
  12   16   39.69  &--12  03    07.2   & 22.00 & 0.5437 	 &    12   16   46.67  &--11  59   37.8  & 21.77 & 0.6669  \\ 
  12   16   39.88  &--11  58   17.0  & 20.55  & 0.2727 	 &    12   16   46.83  &--12  02    22.6  & 21.39 & 0.7987  \\ 
  12   16   39.91  &--11  59   52.9  & 22.81 & 0.7416 	 &    12   16   46.90  &--12  01    23.9 & 22.40 & 0.7910* \\ 
  12   16   39.96  &--11  58   48.1  & 20.50   & 0.2329 	 &    12   16   46.97  &--11  59   26.7  & 21.70 & 0.7971  \\ 
  12   16   40.05  &--12  02    35.2  & 21.12  & 0.8022 	 &    12   16   47.61  &--12  02    28.0  & 20.41 & 0.5434  \\ 
  12   16   40.19  &--12  01    59.3  & 19.34  & 0.3463 	 &    12   16   48.00  &--12  00    22.0  & 21.74  & 0.7860  \\ 
  12   16   40.27  &--12  02    02.9   & 22.18 & 0.7976 	 &    12   16   48.18  &--12  03    18.6  & 22.73 & 0.8039  \\ 
  12   16   40.27  &--11  58   19.8  & 22.90 & 0.6655 	 &    12   16   48.42  &--11  59   10.3  & 19.58 & 0.2735  \\ 
  12   16   40.32  &--11  58   25.4  & 22.95 & 0.2733 	 &    12   16   48.84  &--11  58   30.7  & 21.90 & 0.1504  \\ 
  12   16   40.33  &--12  02    01.4   & 21.41 & 0.7972 	 &    12   16   48.92  &--12  01    23.8 & 22.36 & 0.7940* \\ 
  12   16   40.35  &--11  58   27.7  & 19.91 & 0.2739 	 &    12   16   48.93  &--11  58   57.9  & 22.10 & 1.0742  \\ 
  12   16   40.70   &--12  03    44.0  & 21.00 & 0.7930 	 &    12   16   48.96  &--12  00    09.1   & 21.65 & 0.7863  \\ 
  12   16   40.91  &--12  02    48.8  & 22.94 & 0.9480 	 &    12   16   49.03  &--12  01    42.6  & 22.41 & 0.8000* \\ 
  12   16   41.58  &--11  58   46.4  & 22.93 & 0.8644 	 &    12   16   49.03  &--12  01    53.1  & 22.01 & 0.7998  \\ 
  12   16   41.62  &--11  59   30.8  & 21.94 & 1.0771 	 &    12   16   49.43  &--11  59   16.5  & 22.38  & 0.4082  \\ 
  12   16   41.70   &--12  03   05.4   & 21.36 & 0.8012 	 &    12   16   49.77  &--12  01    35.8  & 21.87 & 0.7882  \\ 
  12   16   41.75  &--12  00    44.9  & 21.66 & 0.7967 	 &    12   16   49.78  &--11  58   34.4  & 23.02 & 0.7885  \\ 
  12   16   41.91  &--12  02    44.0  & 22.10 & 0.8028 	 &    12   16   49.80  &--12  01    39.2  & 21.21 & 0.7965  \\ 
  12   16   42.03  &--12  01    50.9  & 20.62 & 0.7941 	 &    12   16   49.97  &--12  01    10.6  & 21.46 & 0.6980* \\ 
  12   16   42.26 &--12  01    57.2 & 22.95 & 0.7950*	 &    12   16   50.20  &--12  00    03.8   & 22.17 & 0.6660  \\ 
  12   16   42.44  &--12  02    34.8  & 20.65 & 0.2631 	 &    12   16   50.29  &--11  59   59.4  & 22.95 & 0.7906  \\ 
  12   16   42.80   &--12  03    39.5  & 21.33 & 0.7955 	 &    12   16   50.36  &--12  00    12.0  & 22.57  & 0.9312  \\ 
  12   16   42.95  &--11  59   53.6  & 22.01 & 0.7951 	 &    12   16   50.42  &--12  00    48.0  & 21.92 & 0.7886  \\ 
  12   16   43.05  &--11  59   36.5  & 22.18 & 0.2760 	 &    12   16   50.81  &--11  57   57.6  & 21.18  & 0.6501  \\ 
  12   16   43.11  &--11  58   11.3  & 22.50 & 1.0579 	 &    12   16   50.87  &--12  02    05.7  & 20.93 & 0.7960* \\ 
  12   16   43.18  &--12  02    40.7  & 22.46 & 0.3194 	 &    12   16   51.36  &--12  00    31.3  & 22.90 & 0.7841  \\ 
  12   16   43.18  &--11  59   44.4  & 22.54 & 0.7948 	 &    12   16   51.57  &--12  01    30.6  & 22.04 & 0.7220  \\ 
  12   16   43.37  &--12  02    12.8  & 22.05 & 0.7839 	 &    12   16   52.20  &--12  02    26.1  & 20.92 & 0.4062  \\ 
  12   16   43.53  &--12  03    50.2  & 21.31 & 0.6693 	 &    12   16   52.21  &--12  00    59.5  & 22.28 & 0.7882  \\ 
  12   16   43.76  &--12  02    15.5  & 22.37 & 0.8028 	 &    12   16   52.33  &--12  00    22.4  & 22.39 & 0.7583  \\ 
  12   16   43.77  &--11  58   15.5  & 22.83 & 0.6558 	 &    12   16   52.65  &--12  02    55.3  & 21.53 & 0.8263  \\ 
  12   16   43.78  &--12  02    11.1  & 21.57 & 0.7913 	 &    12   16   53.08  &--12  00    45.3 & 22.05 & 0.6490* \\ 
  12   16   43.80   &--12  00    53.6  & 21.10 & 0.7945 	 &    12   16   53.24  &--12  01    36.2 & 21.74 & 0.7930* \\ 
  12   16   43.87  &--11  58   42.5  & 22.79 & 0.7956 	 &    12   16   53.28  &--11  58   54.0  & 21.15 & 0.4763  \\ 
  12   16   43.92  &--12  00    23.3  & 22.20 & 0.7831 	 &    12   16   53.39  &--12  01    38.0 & 22.20 & 0.7925* \\ 
  12   16   44.00   &--11  57   51.6  & 21.92 & 0.7917 	 &    12   16   53.70  &--11  59   27.6  & 22.18  & 0.2723  \\ 
  12   16   44.33  &--12  01    38.4  & 21.72 & 0.7861 	 &    12   16   54.14  &--11  57   55.9  & 22.49 & 0.8748  \\ 
  12   16   44.35  &--12  01    42.9  & 21.05 & 0.7918 	 &    12   16   54.43  &--12  01    32.9 & 22.48 & 0.7900* \\ 
  12   16   44.47  &--12  01    53.3  & 20.52 & 0.6703 	 &    12   16   54.76  &--11  57   45.1  & 21.24 & 0.8746  \\ 
  12   16   44.51  &--12  03    35.9  & 18.48 & 0.2344 	 &    12   16   54.81  &--11  58   03.9   & 22.66 & 0.9827  \\ 
  12   16   44.53  &--12  01    07.5   & 23.55 & 0.7934 	 &    12   16   54.96  &--11  58   10.2  & 20.58 & 0.1034  \\ 
  12   16   44.59  &--12  01    08.9   & 21.83 & 0.8001 	 &    12   16   55.26  &--11  59   23.4 & 22.36 & 0.7950* \\ 
  12   16   44.61  &--12  02    35.8  & 21.93 & 0.6698 	 &    12   16   56.23  &--11  59   39.1 & 22.92 & 0.8740* \\ 
  12   16   44.67  &--12  02    33.7  & 21.61  & 0.6708 	 &        	       & 	       & 	& 	  \\ 
\hline									 
\end{tabular}								  
\label{t:spectro}							  
\end{center}
\end{table*}
}

Remaining slits were put on galaxies along the cluster line-of-sight
(los). For a cross check and for comparison, we included 13 galaxies
in this sample with redshifts already available in the
literature. Combined with publicly available data, our final sample
contains 137 galaxies with redshifts, all with $z \leq$1.15 and $I
\leq 24.5$ (the $I$ magnitude distribution of our sample peaks at
$I=22.5$). Coordinates and redshifts for this sample are given in
Table~\ref{t:spectro} (available electronically only).
Among the 13 galaxies in
common between our GMOS measurements and the literature, only one
(with a low S/N in the GMOS data) showed discrepant redshift
measurements (0.6605 in GMOS vs. 0.7220 in the literature). The
discrepancy probably arises from a different identification of a
spectral feature that we attributed to an H$\delta$ absorption line,
while it was attributed to the Ca H line in the literature. Our
redshift measurements for the other 12 galaxies are in very good
agreement with the previous measurements, with a mean difference of
$-0.0003 \pm 0.0013$. This uncertainty is in agreement with the
expected uncertainty of our GMOS measurements. We adopt
390~km~s$^{-1}$ as the average velocity error for these data.

\subsection{X-ray data}

We have downloaded the publicly available XMM-\textit{Newton} observations of LCDC~0504: ID 0143210801, observed in 07/2003, PI D. Zaritsky, and ID 0651770201, observed in 12/2010, PI B. Maughan. Both observations were reprocessed with SAS~12%
\footnote{Science Analysis System from the XMM-Newton team,
  http://xmm.esac.esa.int/sas/current/sas\_news.shtml}$\!$, using the
latest available calibration files. High background (flares) time
intervals were removed with a $\sigma$-clip method using the
light-curve of the 2.0--12.0 keV band.

The final exposure times, after flare subtraction, are, for the 2003
observation: 22.71, 22.34, and 18.36~ks for the MOS1, MOS2, and pn,
respectively. For the 2010 observation we have 59.96, 63.83, and
27.89~ks for the MOS1, MOS2, and pn, respectively.

\begin{figure}
\begin{center}
\includegraphics[width=8.8cm]{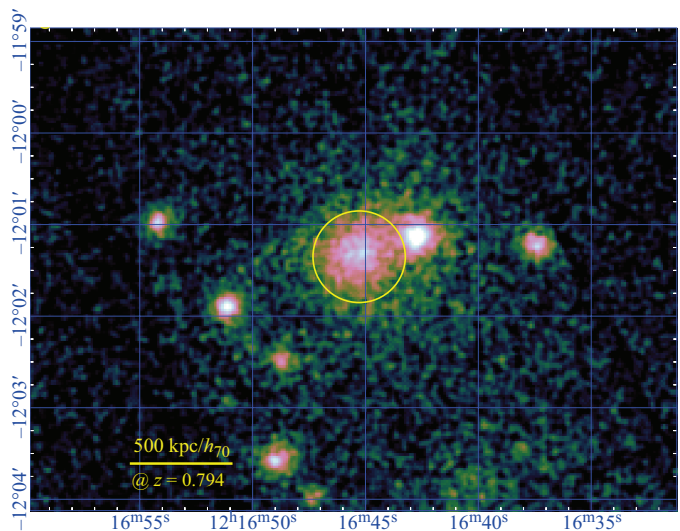}
\end{center}
\caption[]{XMM-Newton image using all available data. The cluster is shown
  inside the yellow central circle of radius equal to
  $30^{\prime\prime}$. The other visible X-ray sources are point
  source AGNs.}
\label{fig:LCDCS0504allxmm0.3_7.0}
\end{figure}

For each observation and detector we have produced exposure-map
corrected images in the 0.3--7.0 keV band. All these images were
merged together using the task imcombine/IRAF, and the result is shown
in Fig.~\ref{fig:LCDCS0504allxmm0.3_7.0}.

\section{Mass profile from strong lensing}
\label{s:sl}
Motivated by the spectroscopy of the blue lensed features in the cluster core
(Section~\ref{ss:spectro}) and after inspection of both the high resolution
HST/ACS image and the ground based $B, R, I$ images, we propose that 
objects 1 through 3 in Table 3 are the result of a
\emph{single} background galaxy at $z=3.0$ is being strongly lensed
by Cl\,1216. We observe two images of this background source.
Each image is resolved into three sub-images, which correspond to different
sub-structures of the background source.
They are labelled systems 1, 2 and 3 (Fig.~\ref{f:multiple}).
We also conjugate two close images, forming system 4. 
Having no redshift information for this system, we let its redshift free during the optimization.

Beginning with this set as constraints, we modeled the cluster mass 
distribution using a dual Pseudo Isothermal Elliptical Mass Distribution 
\citep[dPIE, hereafter;][]{Limousin+05,Eliasdottir+07}. The dPIE model is based
on the Pseudo Isothermal mass profile, characterized by the 3D mass
profile\footnote{This is the total mass enclosed within a radius $r$,
  sometimes written as $M(<r)$ in the literature.}
\begin{equation}
M(r) = 2\,{s\,\sigma_0^2\over G\,(s-a)}\,\left[ s\,\tan^{-1} \left ({r\over s}\right) - a\,\tan^{-1}\left({r\over a}\right) \right] \ ,
\end{equation}
provides a 3D density profile:
\begin{equation}
\rho(r) = {\sigma_0^2 \over 2\pi G}\,{s\,(a+s)\over (r^2+a^2)\,(r^2+s^2)} \ ,
\end{equation}
where $G$ is the gravitational constant, $r$ is the 3D clustercentric
radial distance, $a$ the core radius, $s$ the scale radius, $\sigma_0$
the central velocity dispersion. This profile is not isothermal
(slope $-2$) at all radii but only in the intermediate radial range
$a \leq r \leq s$. This $M(r)$ corresponds to the
projected mass density profile,
\begin{equation}
\Sigma(R)={s \, \sigma_0^2 \over 4 {\rm G} \, (s-a)} \, [(R^2+a^2)^{-1/2}-(R^2+s^2)^{-1/2}] \, ,
\end{equation}
where $R$ is the 2D projected clustercentric radial distance. The
dPIE is obtained by replacing $R$ with
\begin{equation}
\widetilde R^2 = {X^2 \over (1+\epsilon)^2} + {Y^2 \over (1-\epsilon)^2} \, ,
\end{equation}
where the ellipticity is defined as $\epsilon \equiv (A-B)/(A+B)$,
with $A, B$ the semi-major and, respectively, semi-minor axis, and $X,
Y$ are the spatial coordinates along the major, and, respectively,
minor axes. There are 6 free parameters in the dPIE model, the two
coordinates of the cluster center, the ellipticity, the orientation
angle, the velocity dispersion, and the core and scale radii.

We rely on the dPIE model results to identify and search for other gravitational
arcs.  We note, though, that we could not distinguish between the
dPIE model and a NFW model \citep{NFW97}
with our SL analysis, given the uncertainties. However,
we prefer using a dPIE profile since the parameters can be constrained by our 
SL analysis, whereas the NFW profile (in particular the scale radius) is out 
of reach of the SL constraints (but not out of reach of modeling based on
kinematics data, see Sect.~\ref{s:kin}).

We fix the scale radius $s$ to 1 Mpc since it cannot be constrained by
our data.  Furthermore, after some tests, we figured out that the core
radius was constrained to be smaller than $\sim 5\arcsec$,
i.e. smaller than the range where multiply imaged systems are
found. We fix it to 2$\arcsec$.
Note that with this choice of parametrization, the cluster is modelled 
using a mass profile which is close to isothermal.  Given the circular 
aspect of this cluster, we impose its position to be within $\pm$ 5$\arcsec$ 
from the BCG galaxy. Together with the ellipticity and the position angle 
of the mass distribution, this gives 5 parameters to be optimized.  On top of
this smooth component, we include perturbations from the brighest
cluster members located close (i.e. less that $\sim 5\arcsec$) to the
multiply imaged systems. This gives 11 individual galaxies.  Following
earlier works \citep[e.g.][]{Limousin+07b}, we describe these
perturbers using a dPIE profile, whose geometrical parameters
(position, ellipticity, position angle) are set to the one measured
from their light distribution. Their core radius is set to 0, and
their scale radius to 45\,kpc, which describe compact dark matter
haloes, as expected for central cluster galaxies within a tidal
stripping scenario \citep{Limousin+07}. Their velocity dispersion is
scaled with their luminosity (see Limousin et~al. 2007, ApJ for more
details). Therefore, the perturbers are modeled using one extra
parameter.  Using the 8 constraints provided by the multiply imaged
systems, we optimize the mass model in the image plane, using the
\textsc{Lenstool}\footnote{http://www.oamp.fr/cosmology/lenstool/}
software \citep{Jullo+07}.  We find that this simple unimodal
model is able to reproduce accurately the multiply imaged systems,
with an RMS of 0.15$\arcsec$ (image plane).

The mass model predicts a third central image for the strongly lensed
background galaxy at $z=3.0$, predicted to be more than 5 magnitudes
fainter than the main images. System 4 is predicted to be at
$z=2.4\pm0.4$.  Finally, we have not been able to reliably find the
counterimage of the blue feature located at $\alpha=184\fdg18761,
\delta=-12\fdg024721$ (yellow circle on Fig.~\ref{f:multiple}).  One
possibility is that it is singly imaged. In that case, its redshift
should be smaller than 1.35.  

\begin{table}
\begin{center}
\caption{Multiply imaged systems for the SL analysis. 
} 
\begin{tabular}{ccccc}
\hline
ID & R.A.    & Decl.   & $z_{\rm spec}$ & $z_{\rm model}$ \\
   & (J2000) & (J2000) &               &                \\    
\hline
1.1 &184.19186 &--12.01878 & 3.0 & --- \\
1.2 &184.18402 &--12.02390 & 3.0 & --- \\
2.1 &184.18985 &--12.01758 & 3.0 & --- \\
2.2 &184.18683 &--12.02611 & 3.0 & --- \\
3.1 &184.18965 &--12.01752 & 3.0 & --- \\
3.2 &184.18752 &--12.02630 & 3.0 & --- \\
4.1 &184.19216 &--12.01974 & assumed & 2.3$\pm 0.5$ \\
4.2 &184.19171 &--12.01922 & assumed & 2.3$\pm 0.5$ \\
\hline
\end{tabular}
\end{center}
\label{t:multi}
\end{table}

\begin{table}
\begin{center}
\caption{dPIE mass model parameters.}  
\begin{tabular}{ll}
\hline
R.A. (arcsec) & --1.1$\pm$0.8 \\
Decl. (arcsec) & 0.5$_{-2.1}^{+0.9}$ \\
 $\epsilon$ & 0.07$\pm$0.04 \\
 orientation angle (degrees) & 91$\pm$8 \\
$\sigma_0$ (km\,s$^{-1}$) &  839$\pm$14 \\
$a$ (kpc) & [14] \\
$s$ (kpc) & [1\,000] \\
\hline
\end{tabular}
\tablefoot {Coordinates are given in arc-seconds with respect to the cD
  galaxy located at $\alpha =184\fdg18845$, $\delta = -12\fdg021472$.
  Error bars correspond to $1\sigma$ confidence level as inferred from 
  the Monte Carlo Markov Chain optimization.}
\end{center}
\label{t:sl}
\end{table}

\section{Mass profile from X-ray data}
\label{s:x}

\begin{figure}[htb]
\begin{center}
\includegraphics[width=8cm]{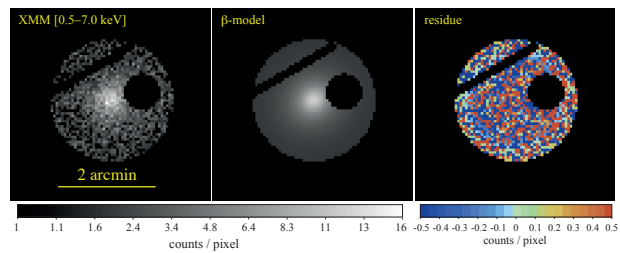}
\end{center}
\caption[]{ 2D surface brightness fit. Left: LCDCS~0504 image in the [0.5-7.0 keV] band with point sources and artefacts (CCD gap) masked out. Middle: best-fit $\beta$-model (see text for details) shown with the same color coding of the original image. Right: residuals, data minus best-fit model. No apparent structure is seen on the residual image.}
\label{fig:2dxrayfit}
\end{figure}

We have produced a surface brightness image of LCDCS~0504 by merging all the individual detectors (MOS1, MOS2, and pn from both 2003 and 2010 exposures) exposure-map corrected images. We then fitted the X-ray surface-brightness profile of
LCDCS~0504 by a 2D, elliptical $\beta$-model\footnote{We use $\beta_X$ for the parameter of the model to distinguish it from the kinematics $\beta$, see eq.~\ref{e:beta}.
\citep{CFF76}, with a flat background added to it:}
\begin{equation}
 I(R) = I_0\,\left[1 + \left({R\over r_c}\right)^2\right]^{1/2-3\beta_X} + B ~,
 \label{eq:beta2D}
\end{equation}
with best-fit values $\beta_X = 0.52 \pm 0.06$ and $r_c = (113 \pm 19)$ kpc, see Fig.~\ref{fig:2dxrayfit}.

 The fit with a $\beta$-model is good, and this suggests the
  cluster is not cool-core, as a cuspy density profile is usually
  observed in cool-core clusters. In non cool-core clusters the
  temperature is usually isothermal inside $r_{500}$. We therefore opt
  for using a single mean temperature for the dynamical modeling. 
  In any case, the data are too sparse to determine such
  a significant non-isothermal nature of the gas so as to change our 
  conclusions. It is not feasible to obtain a meaningful radial
  temperature profile with the $\sim 5800$ net counts (i.e.,
  background subtracted and masking the bright point source to the
  West) resulting from the LCDCS0504 X-ray flux of $(1.0 \pm 0.4) \,
  10^{-13}$ erg~s$^{-1}$~cm$^{-2}$, inside 1 arcmin in the [0.5--10.0]
  keV band.

A spectral analysis was used to compute the central gas density, as
well as its temperature, that was estimated with XSPEC v12, from
HEASARC\footnote{http://heasarc.gsfc.nasa.gov/}. The X-ray spectrum
was extracted within a region of radius 1~arcmin (point sources were
masked) and modeled as an emission from a single temperature plasma
\citep[mekal model;][]{KM93,LOG95}.  We have fitted simultaneously all the spectral data, MOS1, MOS2 and pn from both 2003 and 2010 observations. The photoelectric absorption -- mainly due to Galactic neutral hydrogen -- was computed using the
cross-sections given by \citet{BCM92}, available in XSPEC. Metal
abundances (metallicities) were scaled to the \citet{AG89} solar values.

For the MOS spectra, we restricted our fit to the interval 0.5--7.0~keV,
while for the pn data, we used the 0.7--7.0~keV. We kept the hydrogen
column density fixed for the fit at the Galactic value, $N_{\rm H} =
3.26 \times 10^{22}$~cm$^{-2}$, in the direction of LCDCS~0504
\citep[LAB survey,][]{Kalberla+05}.

\begin{figure}[htb]
\begin{center}
\includegraphics[width=8.8cm]{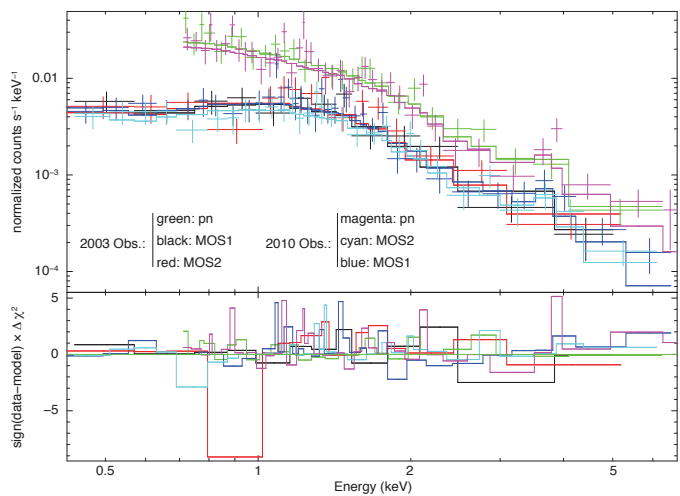}
\end{center}
\caption[]{ Best-fit absorbed MEKAL model. Top: All detectors from both XMM-\textit{Newton} observation are fitted simultaneously, as described in the text. Bottom: Plot of the residual contribution to the $\chi^2$ per energy bin of the best fit spectrum.}
\label{fig:Xspectralfit}
\end{figure}

Our best fit, shown in Fig.~\ref{fig:Xspectralfit} had a reduced $\chi^2 = 0.867$ for 491 degrees of freedom
with the following free parameters: $kT = 5.1^{+0.66}_{-0.55}$~keV and
$Z = 0.23^{+0.17}_{-0.15} Z_\odot$. Inside a radius of 1~arcmin the
fitted spectral model implies a bolometric X-ray luminosity of $L_X =
(2.90 \pm 0.18) \times 10^{44}$~erg~s$^{-1}$. Our spectral fit agrees
quite well with the thorough analysis done by \citet{Johnson+06}, who
used only the first, shallower XMM observation.

Assuming an isothermal plasma, the 3D deprojection of Eq.~(\ref{eq:beta2D}) is: 
\begin{equation}
n(r) = n_0 \, \left[1 + \left({r\over r_c}\right)^2\right]^{-3\beta_X/2} \ ,
\label{eq:ngasbeta}
\end{equation}
where $n_0$ is the central particle number density in units of
cm$^{-3}$ and the radii $r$ and $r_c$ are given in kpc. The central
density was obtained by normalizing the X-ray flux measured with the
expected bremsstrahlung flux from an isothermal $\beta$-model
distribution. From the spectral analysis we obtain $n_0 = (6.5 \pm
0.7) \times 10^{-3}$~cm$^{-3}$.

The total mass profile, assuming isothermal hydrostatic equilibrium
and a spherical $\beta$-model, is given by:
\begin{equation}
 M(r) = 6.68 \times 10^{10}\, \frac{\beta_X kT}{\mu r_c^2}\,
{r^3\over 1 + {r^2}/{r_c^2} }\,
M_\odot \, ,
 \label{eq:mdynx}
\end{equation}
where $\mu = 0.6$, $kT$ is in keV, while $r$ and $r_c$ are in kpc. 
The values of $\rvir$ and $\rs$ corresponding to this mass profile are
given in Table~\ref{t:dyn}. Note that also in this case, as for the
SL determination, the value of $\rvir$ is based on an extrapolation
of the mass profile beyond the region where it is constrained.
The total density profile corresponding to the mass profile of
equation~(\ref{eq:mdynx}) is
\begin{equation}
\rho(r) \propto {r^2+3 r_c^2\over (r^2+r_c^2)^2} \ .
\label{eq:rhox}
\end{equation}

\section{Mass profile from kinematics}
\label{s:kin}

Here, we use the projected phase space distribution of cluster galaxies
to constrain the mass distribution of the cluster.  We adopted three
methods of deriving a mass model based solely on the optical data, two
based on the Jeans equation \citep[e.g.][]{BT87}, and one based on the
Caustic method \citep{DG97,Diaferio99}. The methods based on the Jeans
equation are ``Dispersion-Kurtosis'' \citep[][DK hereafter]{LM03}, and
``MAMPOSSt'' \citep{MBB13}. All three methods assume spherical
symmetry.

The DK method performs a simultaneous best-fit of the parameters of a
model mass profile, $M(r)$, and of a model velocity anisotropy
profile,
\begin{equation}
\beta(r) = 1 - {\sigma_\theta^2(r) + \sigma_\phi^2(r)  \over
  2\,\sigma_r^2(r)} = 1 - {\sigma_\theta^2(r) \over \sigma_r^2\rm{(r)}}  
\label{e:beta}
\end{equation}
where $\sigma_\theta, \sigma_\phi$ are the two tangential components,
and $\sigma_r$ the radial component, of the velocity dispersion, and
the last equivalence is obtained in the case of spherical symmetry.
The fit is done by minimizing the summed $\chi^2$ of the fits to the
binned line-of-sight velocity dispersion profile,
$\slos(R)$, and to the binned line-of-sight kurtosis profile, $K(R)$,
corrected for the known statistical bias using the expression in \citet{decarlo97}.
Using these two profiles rather than just one allows to partially
break the degeneracy between the $M(r)$ and $\beta(r)$
parameters. A limitation of this method is that it assumes that
$\beta(r)$ is constant with radius.

The MAMPOSSt method, like the DK method, determines the best-fit
parameters of model $M(r)$ and $\beta(r)$, but unlike the DK method it
requires no binning of the observables, since it performs a maximum
likelihood fit of the full projected phase space distribution of
cluster members. Unlike\footnote{this is no longer the case with Richardson \& Fairbairn 2013} the DK method, it has no limitation on the
choice of the $\beta(r)$ model. It must however assume a shape for the
3D velocity distribution, and this is taken to be Gaussian in our
analysis.

Both the DK and MAMPOSSt methods assume the cluster to be in dynamical
equilibrium, so their domain of application is limited to the virial
region of the cluster. The Caustic method drops this requirement, and
therefore can be used to determine $M(r)$ also outside the virial
region. However, the Caustic method is less accurate than DK and
MAMPOSSt near the center, and tends to overestimate $M(r)$ at small
radii \citep{Serra+11}.  The Caustic method determines the cluster
mass profile non-parametrically, from the velocity amplitude of the
caustics in projected phase space, but it must assume knowledge of
$\beta(r)$.

\subsection{Cluster membership}
\label{ss:members}
Identification of the cluster members is required in the three
methods,  there are several methods to identify real cluster members
(e.g. \citealp{Wojtak+07,MBB13}).  We applied two of them here to estimate the
uncertainty in the derived results. We used the method of
\citet{dHK96} and the `Clean' method of \citet{MBB13}. We selected
these 2 approaches out of the several discussed as the former was
shown by \citet{Wojtak+07} to perform marginally better than many
other techniques, and the latter is a new method based on the
analysis of the internal dynamics of cluster-sized halos in numerical
simulations \citep{MBM10}.

\begin{figure}
\begin{center}
\begin{minipage}{0.5\textwidth}
\resizebox{\hsize}{!}{\includegraphics{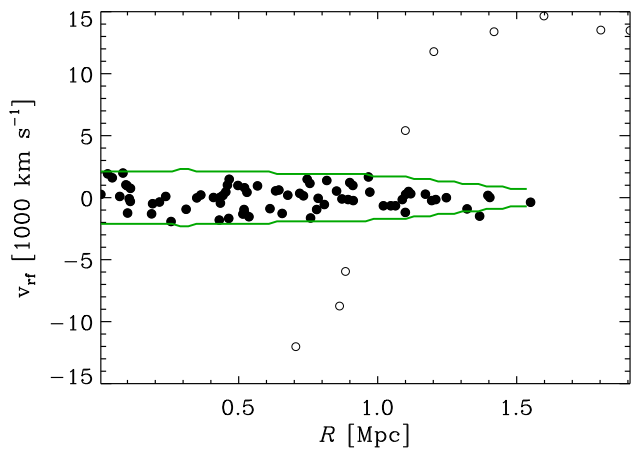}}
\end{minipage}
\end{center}
\caption{The projected phase space distribution of galaxies with
  redshifts in the cluster region.  Selected cluster members are shown
  as filled dots. The chosen caustic in the Caustic method is shown in
  green.}
\label{f:rvcau}
\end{figure}

Both methods identify real cluster members on the basis of their
location in projected phase space\footnote{\label{fn:gal}For each
  galaxy in the cluster, $R$ is the projected cluster-centric distance
  from the cD galaxy, and ${v_{\rm rf}}$ is the rest-frame velocity
  \mbox{${v_{\rm rf}} \equiv ({v-v_{\rm cl}})/(1+{v_{\rm cl}/c})$},
  where ${v_{\rm cl}}$ is the mean velocity of the cluster. This is
  re-defined at each new iteration of the membership selection, until
  convergence. The cluster center is defined to be the position of the
  cD galaxy.}, $R, {v}_{\rm rf}$. The two methods identify the
same galaxies as members of LCDCS~0504, 75 in total (see
Fig.~\ref{f:rvcau}). Based on this sample we estimate the cluster
velocity dispersion $\slos=974_{-76}^{+83}$ km~s$^{-1}$ \citep[biweight
  estimate, see][]{BFG90}.

\subsection{Galaxy number density profile}
\label{ss:nr}
In both the DK and MAMPOSSt methods, the number density profile of the
tracers of the gravitational potential, $n(r)$, needs to be
estimated.  This determination of $n(r)$ is the only occurrence in our
dynamical analysis where completeness, or correction for
incompleteness, is necessary. Since our spectroscopic sample is not
complete, we use the 100\% complete sample of galaxies with magnitude
$F814 \leq 24$ and measured photometric redshifts, $z_{\rm p}$, for the
determination of $n(r)$.

\begin{figure}
\begin{center}
\begin{minipage}{0.5\textwidth}
\resizebox{\hsize}{!}{\includegraphics{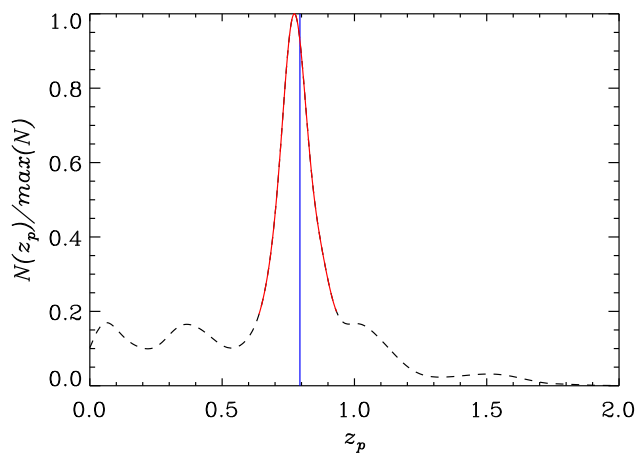}}
\end{minipage}
\end{center}
\caption{The adaptive-kernel smoothed distribution of photometric
redshifts for galaxies in the cluster region. The vertical (blue) line
shows the average cluster redshift. The solid (red) curve shows
the selected $z_{\rm p}$ range for the sample used for the determination
of $n(r)$.}
\label{f:zp}
\end{figure}

Our photometric observations fully cover the cluster only out to $\sim
2$ arcmin from the adopted cluster center, the cD galaxy.  Beyond this
radius we estimate the radial geometrical completeness, $C_g(R)$, as
the fractions of circular annuli covered by our observations. $C_g(R)$
drops below 50\% beyond 3.3 arcmin.

We select the $z_{\rm p}$-range for defining cluster membership as
follows. We smooth the $z_{\rm p}$ distribution by an adaptive kernel
technique with a kernel size of 0.045, i.e. half the value of the
  typical uncertainty on the photometric redshifts.  Larger values of
  the kernel size would lead to un-necessary over-smoothing of the
  $z_{\rm p}$ distribution, while smaller values are likely to
  emphasize noise-related features. We identify the main peak of
  the $z_{\rm p}$ distribution closest to the mean cluster
redshift. We define the extremes of this peak in $z_{\rm p}$ in such a
way as to avoid contamination from other peaks in the distribution,
$0.64 \leq z_{\rm p} \leq 0.93$ (see Fig.~\ref{f:zp}).

\begin{figure}
\begin{center}
\begin{minipage}{0.5\textwidth}
\resizebox{\hsize}{!}{\includegraphics{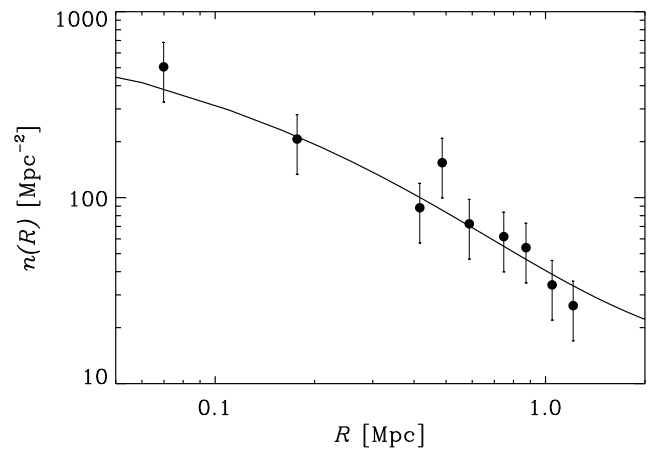}}
\end{minipage}
\end{center}
\caption{The projected number density profile of $z_{\rm p}$-selected
  cluster members (points with 1$\sigma$ error bars) and the best-fit
  model (projected-NFW + constant density background; solid curve).
  The reduced $\chi^2$ of the fit is 0.8.}
\label{f:nr}
\end{figure}

We perform a maximum-likelihood fit of the spatial distribution of the 375
galaxies in the selected $z_{\rm p}$-range, weighting each galaxy by
$C_g(R_i)^{-1}$, where $R_i$ is the radial position of galaxy $i$, to
account for geometrical incompleteness. We limit the fit of the number
density profile to the radii where $C_g \geq 0.5$.  The fitted model
is NFW in projection \citep{Bartelmann96,LM01} to which we add a
constant density background to account for interlopers in our $z_{\rm p}$
selection.  Of the two free parameters of the NFW model, we are only
interested in the scale radius of the galaxy number density, $\rtr$,
because the other parameter, that sets the normalization of $n(r)$,
cancels out in the Jeans equation. We find $\rtr=0.27_{-0.15}^{+0.44}$
Mpc, and a background density corresponding to 38\% background
contamination in our $z_{\rm p}$-selected sample.

Since the uncertainties on $\rtr$ are very large, we also consider an
alternative estimate, based on the spectroscopic sample of cluster
members (see Sect.~\ref{ss:members}). The radial geometrical
completeness, $C_g(R)$, is the same for this sample as for the
$z_{\rm p}$-selected sample. In addition, the spectroscopic sample
suffers from radially-dependent completeness because the fraction of
galaxies with measured redshifts is higher near the cluster center. We
evaluate this spectroscopic completeness, $C_s(R)$, as the ratio
of the number of galaxies with measured redshifts to the total number
of galaxies (134 and 713 in total)
in radial bins, down to $F814 \leq 23$. We find $C_s(R)=0.22$
outside the central bin, i.e. at $R \geq 0.11$ arcmin, and 
$C_s(R)=0.50$ inside this bin. We then run a maximum-likelihood
fit of the spatial distribution of spectroscopic members weighting
each galaxy by $[C_g(R_i) \times C_s(R_i)]^{-1} $. We find
$\rtr=0.48_{-0.24}^{+0.46}$ Mpc, and a background density
corresponding to 4\% background contamination. The background
contamination, which is much lower than for the $z_{\rm p}$-selected sample,
as expected. The $\rtr$ value is consistent within the (large) error bars
with that obtained using the $z_{\rm p}$-selected sample.

In the dynamical analysis with the DK and MAMPOSSt methods, we will
 use both estimates of $\rtr$, to understand how much our
limited knowledge of $\rtr$ affects our estimate of the cluster mass
profile. The knowledge of $\rtr$ is not required for the dynamical
analysis with the Caustic method.

\subsection{Results}
\label{ss:dynres}
For both the DK and MAMPOSSt methods we use the NFW 
model for $M(r)$ \citep{NFW97}, 
\begin{equation}
M(r)=\mvir {\ln(1+r/\rs)-r/\rs \, (1+r/\rs)^{-1} \over \ln(1+\cvir)-\cvir/(1+\cvir)},
\end{equation}
where $\cvir \equiv \rvir/\rs$ is the mass profile concentration.  The
model has two free parameters, the virial mass and concentration, or,
equivalently, the virial and scale radii $\rvir$ and $\rs$. Note that
the total mass density scale-length is different from the scale-radius
of the galaxy number density profile, i.e. $\rs \neq \rtr$ (Sect.
\ref{ss:nr}), since we allow the ditribution of the total mass and
that of the galaxies to be different in our analysis.  For the Caustic
technique, for the sake of comparison with the other two methods, we
also fit a NFW model to the mass density profile determined from
differentiation of the non-parametrically determined mass profile.

The MAMPOSSt method is the only one among the three where there is
complete freedom in the choice of $\beta(r)$. We use a
simplified version of the model of \citet{Tiret+07},
$\beta(r)=\beta_{\infty} \, r/(r+\rs)$, where
$\beta_{\infty}$ is the asymptotic value of the anisotropy reached at
large radii, and $\rs$ is the scale radius of the NFW mass
density distribution. This model was shown by \citet{MBM10} to provide
a good fit to cluster-mass halos extracted from cosmological 
numerical simulations. In this model, galaxy orbits are isotropic near
the cluster center and become increasingly radially anisotropic outside.

In the Caustic technique, we use Gaussian adaptive kernels for the
density estimation in projected phase space, with an initial kernel
size equal to the optimal kernel size of \citet{Silverman86}. Before
the density estimation, we scale the velocity coordinates such that the
scaled velocity dispersion is the same as the dispersion in the radial
coordinates.  In the equation that connects $M(r)$ to the Caustic
amplitude \citep[eq. 13 in][]{Diaferio99},
we adopt either ${\cal F}_{\beta}=0.5$ as recommended by
\citet{Diaferio99} and \citet{GDRS13}, or ${\cal F}_{\beta}=0.7$ as
recommended by \citet{Serra+11}.  For the estimation of the $M(r)$ error
we adopt the recipe of \citet{Diaferio99}. \citet{Serra+11} have found
that these error estimates correspond to 50\% confidence levels; we
therefore scale them up by a factor of 1.4 to have $\sim 1\sigma$
level error estimates.  The chosen caustic is displayed in
Fig.~\ref{f:rvcau}.

The domain of application of the DK and MAMPOSSt methods is the virial
region. Since almost all our cluster members are in the virial region,
we only exclude the very central region, 25 kpc, where the
gravitational potential is likely to be dominated by the cD and therefore
unlikely to follow a purely NFW profile.


\begin{table}
\begin{center}
\caption{\label{t:dyn}Best-fit $M(r)$ and $\beta$ parameters from kinematics.}
\begin{tabular}{lllc}
\hline
\\
Method & $\rvir$ & $\rs$  & Velocity  \\
& [Mpc] & [Mpc] & anisotropy \\
\\
\hline
\\
DK ($p$)       & $1.28_{-0.06}^{+0.16}$ & $0.05_{-0.03}^{+0.26}$ & $-3_{-1}^{+3}$ \\
\\
DK ($s$)       & $1.28_{-0.10}^{+0.12}$ & $0.05_{-0.05}^{+0.35}$ & $-3_{-2}^{+3}$  \\
\\
MAMPOSSt ($p$) & $1.32_{-0.09}^{+0.14}$ & $0.16_{-0.10}^{+0.34}$ & $0.5_{-0.2}^{+0.4}$ \\
\\
MAMPOSSt ($s$) & $1.28_{-0.12}^{+0.10}$ & $0.19_{-0.12}^{+0.43}$ & $0.5_{-0.2}^{+0.4}$ \\
\\
Caustic,  ${\cal F}_{\beta}=0.5$ & $1.27_{-0.20}^{+0.16}$ & $0.14_{-0.04}^{+0.06}$ & -- \\
\\
Caustic,  ${\cal F}_{\beta}=0.7$  & $1.42_{-0.22}^{+0.18}$ & $0.16_{-0.04}^{+0.04}$ & -- \\
\\
Weighted average & $1.30 \pm 0.05$ & $0.14 \pm 0.07$ & -- \\ 
\\
\hline
\end{tabular}
\tablefoot{The DK and MAMPOSSt estimates are obtained using the number
  density profile based on the photometric ($p$) or the spectroscopic
  ($s$) samples of cluster members. The velocity anisotropy is the
  constant value of $\beta$ for the DK method, and $\beta_{\infty}$ in
  the simplified Tiret model for the MAMPOSSt method.}
\end{center}
\end{table}

\begin{figure}
\begin{center}
\begin{minipage}{0.5\textwidth}
\resizebox{\hsize}{!}{\includegraphics{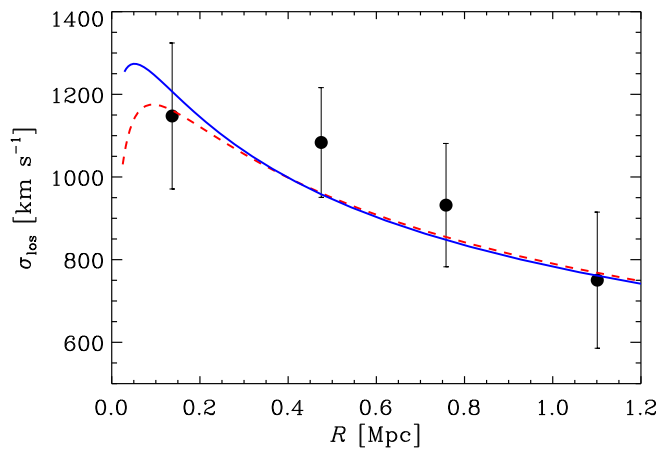}}
\end{minipage}
\end{center}
\caption{The observed line of sight velocity dispersion profile (points with
  1~$\sigma$ error bars) and those predicted by the best-fit NFW
  models, obtained with the DK (dashed red line), and MAMPOSSt (solid
  blue line) methods. Only those solutions obtained using the $\rtr$
  value found with the $z_{\rm p}$ sample of members are shown, for clarity.}
\label{f:vdp}
\end{figure}

\begin{figure}
\begin{center}
\begin{minipage}{0.5\textwidth}
\resizebox{\hsize}{!}{\includegraphics{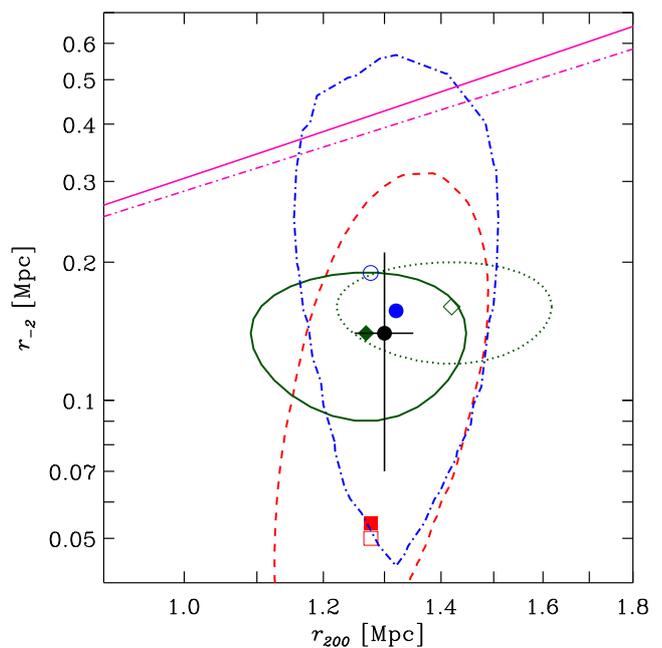}}
\end{minipage}
\end{center}
\caption{The best-fit $M(r)$ NFW parameters from the kinematics
  analyses, within 1~$\sigma$ confidence level contours, obtained with
  the DK (red squares), MAMPOSSt (blue dot and circle), and Caustic
  (green filled and open diamond) methods.  The filled (resp. open)
  symbols are for the solutions obtained using the $\rtr$ value from
  the photometric (resp. spectroscopic) sample of members.  The dashed
  red (resp. dash-dotted blue) contour represents the 1 $\sigma$
  confidence region on the best-fit parameters of the DK
  (resp. MAMPOSSt) method, obtained using the $\rtr$ value from the
  photometric sample of members.  The solid (resp. dotted) green
  contour represents the 1 $\sigma$ confidence region on the best-fit
  parameters for the Caustic method obtained using ${\cal
    F}_{\beta}=0.5$ (resp.  ${\cal F}_{\beta}=0.7$).
  The magenta solid (resp. dash-dotted) inclined line is the
  theoretical predictions for relaxed clusters at the mean redshift of
  LCDCS~0504 from \citet{BHHV13} \citep[resp.][]{DeBoni+13}. 
The black dot with error bars is the weighted average of the DK, MAMPOSSt and
Caustic results. 
}
\label{f:rvrs}
\end{figure}

The results of the dynamical analysis are summarized in
Table~\ref{t:dyn} and displayed in Fig.~\ref{f:rvrs}, where we show
the confidence contour in the $[\rvir, \rs]$ plane.  We also list
  and display a weighted average of the DK, MAMPOSSt, and Caustic
  results. We multiply the formal error on this average by $\sqrt{5}$
  to take into account that the five averaged results are not
  independent. The constraints on the $\beta$ parameters obtained by
the DK and MAMPOSSt methods are very loose, so we do not display them
here.

There is a good agreement between the values of $\rvir$ obtained by
the DK, MAMPOSSt and Caustic methods. The Caustic solution obtained
with ${\cal F}_{\beta}=0.5$ is closer to those from the other two
methods. This would argue in favor of using this value, rather
than ${\cal F}_{\beta}=0.7$, in the Caustic technique, as done
recently by \citet{GDRS13}. However, \citet{Gifford+13} have
  recently suggested using the intermediate value ${\cal F}_\beta$=0.65.
  Moreover, for $\beta=\rm cst$ NFW models, at the half-mass
  radius of $\approx 2\,r_{-2}$, ${\cal F}_\beta=0.5$ corresponds to
  $\beta=-1.1$ while ${\cal F}_\beta=0.7$ corresponds to
  $\beta=0.3$. The very tangential anisotropy for ${\cal F}_\beta=0.5$
  is not what most analysis extract for galaxies in clusters
  \citep[e.g.][and references therein]{Biviano+13}, so the agreement
  of the ${\cal F}_{\beta}=0.5$ solution with those of the DK and
  MAMPOSSt methods may just be fortuitous.

The DK and MAMPOSSt methods pose very weak constraints on $\rs$. \cite{MBB13}
already noted that the determination of the dark matter scale radius is
inefficient, which \cite{SLM04} had previously noted for the concentration
parameter.
The constraints obtained by the Caustic technique are tighter, and
almost independent of the value of ${\cal F}_{\beta}$. The Caustic
technique is able to better constrain the $\rs$ parameter of the mass
distribution than the DK and MAMPOSSt techniques possibly because,
unlike these, it is the only free parameter in the model fit. In fact
$\beta(r)$ is fixed when the value of ${\cal F}_{\beta}$ is assumed,
and $\rvir$ is estimated non-parametrically directly from the Caustic
mass profile. 

The agreement between the MAMPOSSt and DK solutions is also
evident from Fig.~\ref{f:vdp} where we show the projection of the
best-fit DK and MAMPOSSt solutions on the observed line-of-sight
velocity dispersion profile\footnote{We remind the reader that the DK
  best-fit solution is obtained by a simultaneous fit of both the
  observed velocity dispersion profile and the observed kurtosis
  profile, while the MAMPOSSt best-fit solution is obtained by a fit
  of the full line-of-sight velocity distribution. Fig.~\ref{f:vdp}
  is just a way of presenting the best-fit models.}. 

In Fig.~\ref{f:rvrs} we also show the theoretical predictions of
\citet{BHHV13} and \citet{DeBoni+13} for the concentration-mass
relation of relaxed clusters at the redshift of LCDCS~0504, converted
in the $\rvir$-$\rs$ plane. Both theoretical predictions predict
too high a concentration for a cluster of the mass and at the redshift
of LCDCS~0504.  Since the prediction of \citet{DeBoni+13} is based on
hydrodynamic simulations, while that of \citet{BHHV13} originates from
DM-only simulations, the discrepancy between theoretical
  predictions and observation cannot be explained by baryonic
  processes affecting the cluster dynamical structure.

\section{Comparing the different mass profile determinations} 
\label{s:cmp}

  The methods based on SL, X-ray, and kinematics to determine the
  cluster mass profile have different sensitivities on different
  scales. It would therefore be misleading to either extrapolate the
  SL and X-ray mass estimates to $\rvir$ to compare with the result
  from kinematics, or to restrict the spectroscopic data-sample to a
  smaller region to directly infer $r_{2500}$ or $r_{500}$ from the
  kinematics analysis, with loss of statistics. Rather than comparing
  the mass profile parameters, a more appropriate comparison is that
between the different mass profiles themselves, in the regions
  where they overlap.

\begin{figure}
\begin{center}
\begin{minipage}{0.5\textwidth}
\resizebox{\hsize}{!}{\includegraphics{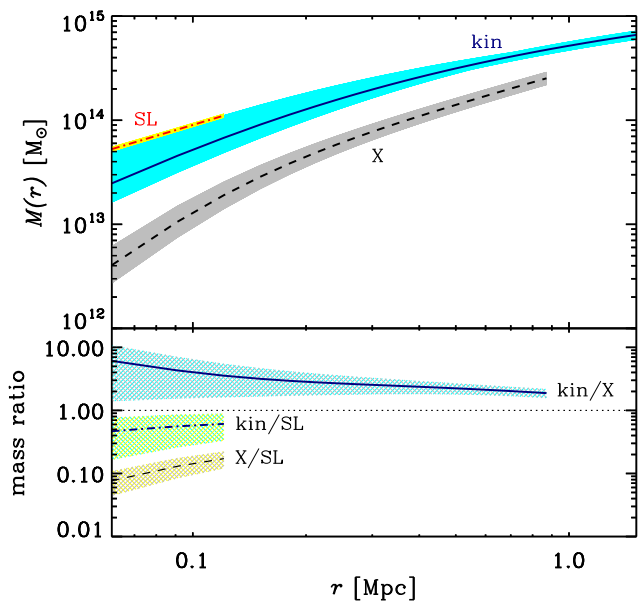}}
\end{minipage}
\end{center}
\caption{{\em Top panel:} The mass profiles and their 1~$\sigma$
  confidence regions obtained from the SL (red dashed line and yellow
region), X-ray (black dashed line and grey region), and kinematics
(blue solid line and cyan region) analyses. {\em Bottom panel:}
the ratios of the three mass profiles and their 1~$\sigma$ confidence
regions. Solid blue line and grey-cyan region: ratio of the kinematics to
X-ray mass profiles. Dashed-dotted blue line and green region: ratio 
of the kinematics to SL mass profiles. Dashed black line and orange
region: ratio of the X-ray to SL mass profiles. In both panels the profiles
are shown in the radial range where they are constrained by the data.}
\label{f:mr}
\end{figure}

The three $M(r)$ from the SL, X-ray, and kinematics analyses are shown
in the top panel of Fig.~\ref{f:mr}, their ratios are shown in the
bottom panel of the same figure\footnote{To compare the mass
  distribution obtained with the SL analyses with the others, we take
  the spherical approximation also for the SL method. In practice we
  force to zero the ellipticity parameter $\epsilon$ of the SL
  model.}. The SL and kinematics $M(r)$ are in agreement within the
errors. The significant difference in the $\rvir$ values of these two
profiles is therefore due to the uncertain extrapolation of the SL
$M(r)$, which is considerably flatter than the kinematics $M(r)$.
On the other hand, within inner 100 kpc, the $M(r)$ obtained by the 
X-ray analysis
is significantly below both the SL and the kinematics mass profiles.
In this case, the discrepancy is real and cannot be attributed to
extrapolation uncertainties.

We will discuss the possible origin of the differences between the mass
profiles in Sect.~\ref{s:summ}.

\section{The gas mass fraction}
\label{s:bary}
We compute the intra-cluster gas mass profile with the integral of
the gas density profile of equation~(\ref{eq:ngasbeta})  
over a spherical volume. For the present
cluster we have:
\begin{equation}
 M_{\rm gas}(r) = 1.205 \times 10^8\, n_0 \, r^3 \, {}_2F_1\left(\frac{3}{2},\frac{3 \beta_X}{2},\frac{5}{2}, - \frac{r^2}{r_c^2} \right) M_\odot \, ,
 \label{eq:mgas}
\end{equation}
where $_2F_1(a,b,c,x)$ is the standard hypergeometric function.\footnote{For
  $\beta_X=1/2$, consistent with our fit to the X-ray surface brightness
  profile, a useful approximation to the hypergeometric function of
  equation~(\ref{eq:mgas}) is 
${}_2F_1\left(\frac{3}{2},\frac{3}{4},\frac{5}{2}, - \frac{r^2}{r_c^2}
  \right)
\simeq 6 \left [(x^3/3)^{-\gamma}+(2 x^3/3)^{-\gamma}\right]^{-1/\gamma}$,
where $\gamma = 2^{1/8} \simeq 1.0905$,
which is accurate to better than 2.7\% for all radii
(see \citealp{ML05b}).}

\begin{figure}[htb]
\centering
\includegraphics[width=8.8cm]{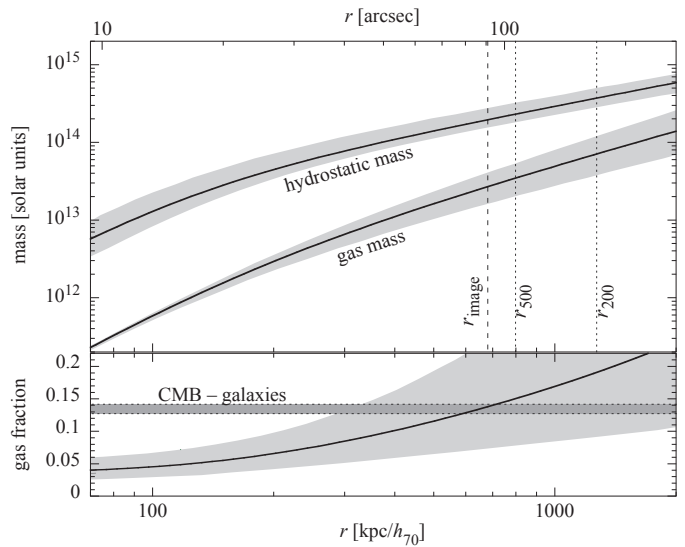}
\caption[]{{\em Top panel}: The intra-cluster gas mass profile
  (lower curve) and the hydrodynamical derived total mass radial
  profiles. The grey shaded regions represent $1 \sigma$ confidence
  levels. Vertical lines indicate $r_{image}$, the limit where the
  cluster is detected with the combined XMM data (using both
  exposures), and $r_{500}$ and $\rvir$, computed from the X-ray
  derived mass profile.  {\em Bottom panel}: the gas mass fraction
  radial profile. As a reference we also show the universal gas
  fraction, as obtained by the cosmic baryon fraction
  $\Omega_{b}/\Omega_{m}$ value from WMAP-9yr \citep{Hinshaw+12}
  and Planck 1st release \citep{PlanckCollaboration+13} (including
  their uncertainties), reduced by 17\%}
\label{fig:xraymass}
\end{figure}

Dividing Eq.~(\ref{eq:mgas}) by Eq.~(\ref{eq:mdynx}) yields the gas
mass fraction, $f_{\rm gas}$. Figure~\ref{fig:xraymass} shows the mass
profiles, gas and total, in the upper panel and the gas fraction
radial profile in the bottom panel. The cluster gas mass fraction
increases with $r$, as seen in most clusters \citep[see,
  e.g.,][]{BS06,Allen+08,FHHR09}. At $r>300$ kpc, the gas mass
  fraction reaches a value that is consistent with the cosmic gas
  fraction, which is $\simeq 83$\% of the cosmic baryon fraction
\citep{FHP98,Hinshaw+12,PlanckCollaboration+13}. 

\begin{figure}[!htb]
\begin{center}
\includegraphics[width=8.8cm]{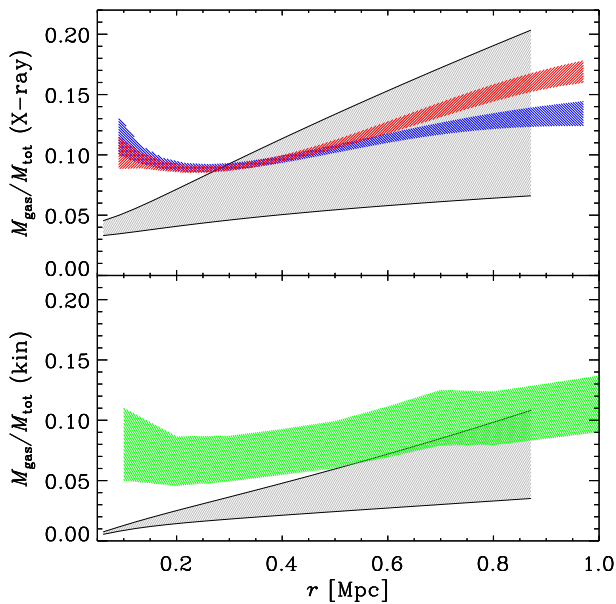}
\end{center}
\caption{{\em Top panel:} The ratio of gas mass to total mass from the
  X-ray analysis. The grey shaded region within solid lines is the 1
  $\sigma$ interval on the observed gas mass fraction of LCDCS~0504.
  The blue and red shaded regions (the blue one below the red one at
  large radii) are the average gas mass fractions for cool-core and
  non-cool-core clusters from \citet{Eckert+13}. {\em Bottom panel:}
  The ratio of gas mass to total mass, the latter derived from the
  kinematics analysis. The grey shaded region within solid lines is
  the 1 $\sigma$ interval on the observed gas mass fraction of
  LCDCS~0504. The green shaded region is the average gas mass fraction
  for nearby clusters from \citet{BS06}.}
\label{f:fgas}
\end{figure}

A comparison with the results of \citet{Eckert+13} shows that the gas
mass fraction profile of LCDCS~0504 is very similar to that of
lower-$z$ clusters, except near the cluster center, where it is
significantly below. This is shown in the top panel of
Fig.~\ref{f:fgas} where we plot the results of \citet{Eckert+13}
for the average gas mass fraction of cool-core and non-cool-core
clusters, together with our results, based in both cases on the
total mass determined from X-ray analysis. If we instead use
the total mass determined from kinematics, we can compare our
result with that of \citet{BS06}. This comparison is shown in
the bottom panel of Fig.~\ref{f:fgas}. In this case the gas fraction of
LCDCS~0504 appears to lie below that for a sample of nearby clusters
at almost all radii.

\begin{figure}[!htb]
\begin{center}
\begin{minipage}{0.5\textwidth}
\resizebox{\hsize}{!}{\includegraphics{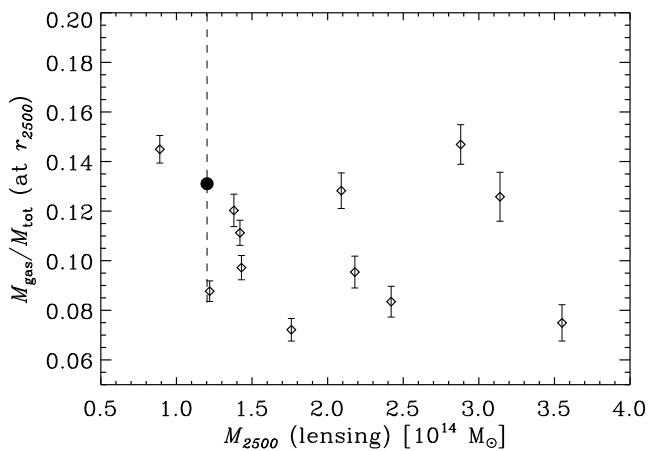}}
\end{minipage}
\end{center}
\caption{The ratio of gas mass to total mass determined from lensing
  analyses for the clusters of \citet{Zhang+10} (diamonds) and for
  LCDCS~0504 (dot). Error bars are 1 $\sigma$.}
\label{f:fgaslens}
\end{figure}

Finally, we compare the LCDCS~0504 gas mass fraction computed using
the total mass derived from our lensing analysis (see
Sect.~\ref{s:sl}) with those of \citet{Zhang+10}, also derived using
total mass estimates from lensing, except that in their case it was
the weak, not the strong lensing effect that was used. The comparison
is shown in Fig.~\ref{f:fgaslens}, where we plot the gas mass fractions
as a function of the cluster mass, both determined at $r_{2500}$.
This is the smallest radius at which \citet{Zhang+10} have given
their determinations and still it is beyond the region where
SL is detected in LCDCS~0504, $r_{2500}=0.49$ Mpc for the SL $M(r)$.
From Fig.~\ref{f:fgaslens} we see that the gas mass fraction of
LCDCS~0504 at this radius is not anomalous. This is consistent
with the conclusions we obtained using the X-ray- and 
kinematics-determined total masses (Fig.~\ref{f:fgas}), LCDCS~0504
shows an anomalous (low) gas mass fraction only at small radii.

We will discuss in the next Section the possible origin of this
central gas fraction deficiency.

\section{Discussion and conclusions}
\label{s:summ}
We have analyzed the mass profile $M(r)$ of a $z \approx 0.8$ cluster
with the SL technique, using the X-ray emission from the intra-cluster
hot gas, and using galaxies as tracers of the gravitational
potential. The different determinations of the cluster $M(r)$ disagree, 
especially in the inner regions. 
The SL $M(r)$ is slightly above but still consistent with the kinematic determination, but both are significantly above the X-ray $M(r)$ determination.

This discrepancy is unlikely to be caused by an unrelaxed
dynamical status of the cluster. This could cause an overestimate of
the cluster velocity dispersion and hence the cluster mass estimate
from kinematics \citep[see, e.g.,][]{Biviano+06} and an incomplete
thermalization of the intra-cluster gas, leading to an underestimate
of the cluster mass estimates from X-ray \citep[e.g.][]{Rasia+06},
but it would not affect the lensing mass estimate. Moreover, an
  unrelaxed dynamical status is not supported by the analyses of
substructures by \citet{Guennou+13}.
In that paper, we have
used the \citet[][SG hereafter]{SG96} hierarchical method for the
detection of substructures in the distribution of galaxies and
searched for substructures in the X-ray data (described in
Sect.~\ref{ss:imaging}), by analysing the residuals of the subtraction
of a symmetric elliptical $\beta$-model from the X-ray image (see
Guennou et al. 2013 for details). Seven substructures were detected by
the SG technique, all with masses below 10\% of the total cluster
mass. Of these, only one was also detected in X-rays, with an X-ray
luminosity of $\approx 8 \%$ the total cluster X-ray luminosity.  This
analysis indicates that any major perturbation of the LCDCS~0504
dynamical status must thus have occurred sufficiently long ago for the
remnants of the merging groups to have disappeared.

  Another interesting possibility is that we see the cluster with
  its major axis along the line-of-sight. This is suggested by the
  circularly symmetric SL configuration and by the small ellipticity
  of the cD galaxy, since the elongation of cD galaxies generally
  reflects those of their host clusters \citep[e.g.][]{RK87,KASL02}
  (see Fig.~\ref{f:multiple}). It has been shown both on numerical
  simulations \citep{KE05} and observationally \citep{Wojtak13}, that
  clusters are prolate not only in position space but also in velocity
  space, and the major axes of the spatial and velocity distributions
  are aligned. The orientation of the cluster with the major axis
  along the line-of-sight then results in an overestimate of the
  cluster mass estimatd from SL and velocity dispersion.  According to
  \cite{Wojtak13}, the mean ratio of the velocity dispersions along
  the minor and major axes of a cluster is $\simeq 0.78$. This implies
  a ratio of the velocity dispersion along the major axis to the mean
  cluster velocity dispersion of 1.16, i.e. a 32\% mass overestimate
  at a given radius.  This is still not sufficient to remove the
  systematic difference between the mass profile derived from
  kinematics and that derived by the X-ray analysis.

  The alignment effect just discussed could also induce an
  overestimate of mass profile concentration value. This could
explain the disagreement we find with the theoretical predictions 
  of \citet{BHHV13} and \citet{DeBoni+13} (see
  Fig.~\ref{f:rvrs}). 

Whatever the cause for the X-ray vs. kinematics and SL $M(r)$
discrepancy, substantial systematic underestimates of cluster masses
by the X-ray methodology could be interesting for cosmology, as it
could alleviate the tension between the $\sigma_8$ values found by the
Planck collaboration using the Cosmic Microwave Background
power-spectrum on one hand and cluster counts obtained by the
Sunyaev-Zeldovich effect (using X-ray masses as mass calibrators
\citealp{Planck13}): If X-ray masses are underestimated at given SZ
signal, this means the distribution of SZ counts above a given mass
threshold is underestimated, meaning that $\Omega_{\rm m}$
($\sigma_8$) is underestimated (overestimated), which would bring the
best-fit value more in line with the CMB value.

Another intriguing result of our analysis is the discovery that the
gas mass fraction is anomalously low near the center of the LCDCS~0504
cluster. Given the relaxed, symmetric morphology of the X-ray emission
(see Fig.~\ref{fig:LCDCS0504allxmm0.3_7.0}), it is unlikely that this
anomaly could be attributed to the effects of a major merger
displacing the gas from the center, as in the case of the Bullet
cluster \citep{Barrena+02,Markevitch+02}. Alternatively, the gas could
have been ejected by AGN outbursts, while the effects of SNe explosions
should not be significant \citep{CO08,Dubois+13}. \citet{Dubois+13}
predict a 30\% loss in the core due to AGN outflows, which is not to
far from our observed deficiency (with respect to the average of
other clusters) of \mbox{$\simeq 60$}\% (see Fig.~\ref{f:fgas}), given the large
observational uncertainties.

The main issue with the AGN hypothesis is that there is no evidence of
a radio source in the NVSS catalog or in the X-rays as there is no
detectable point source at the location of the cD, although there is a
hint of a cool core.  Also, we have no evidence of broad lines in the
optical spectrum of the cD. All this lack of evidence does, however,
tell us, is that the assumed AGN activity have subsided long enough
ago so that all strong electromagnetic signatures of AGN activity have
now subsided.

In the near future, we plan to extend the dynamical and structural
analysis presented here to clusters with sufficient spectroscopic
information in the full DAFT/FADA cluster set.  Expanding our
data-sets should allow us to determine if the anomalies identified in
LCDCS~0504 are a characteristic of high-$z$ clusters or
not. Hopefully, with a larger sample we will be able to unveil the
hidden systematics causing discrepant determinations of cluster mass
profiles by different methods, and to relate these systematics to the
currently not well understood physics of the intra-cluster baryons.

\begin{acknowledgements}

We wish to express our sincere condolences and grief to the family of
Alain Mazure who unexpectedly passed away during the preparation of
this paper. We wish to thank the anonymous referee for her/his suggestions. 
ML acknowledges the Centre National de la Recherche Scientifique
(CNRS) for its support.  The Dark Cosmology Centre is funded by the
Danish National Research Foundation.  This work has been conducted
using facilities offered by CeSAM (Centre de donn\'eeS Astrophysique
de Marseille -- http://www.lam.fr/cesam/).  FD acknowledges long-term
support from CNES and CAPES/COFECUB program 711/11.  AB acknowledges
the hospitality of the Inst. d'Astroph. de Paris and of the Obs. de la
C\^ote d'Azur.  This research has made use of the NASA/IPAC
Extragalactic Database (NED) which is operated by the Jet Propulsion
Laboratory, California Institute of Technology, under contract with
the National Aeronautics and Space Administration. GBLN and ESC acknowledge 
the support of the Brazilian funding agencies FAPESP and CNPq. Based on 
observations obtained at the Gemini Observatory, which is operated by the 
Association of Universities for Research in Astronomy, Inc., under a 
cooperative agreement with the NSF on behalf of the Gemini partnership: 
the National Science Foundation (United States), the National Research 
Council (Canada), CONICYT (Chile), the Australian Research Council 
(Australia), Minist\'{e}rio da Ci\^{e}ncia, Tecnologia e Inova\c{c}\~{a}o 
(Brazil) and Ministerio de Ciencia, Tecnolog\'{i}a e Innovaci\'{o}n Productiva 
(Argentina). I.M. acknowledges financial support from the Spanish grant
AYA2010-15169 and from the Junta de Andalucia through TIC-114 and the
Excellence Project P08-TIC-03531.

\end{acknowledgements}

\bibliography{master_fin}

\begin{thebibliography}{87}
\expandafter\ifx\csname natexlab\endcsname\relax\def\natexlab#1{#1}\fi

\bibitem[{{Adami} {et~al.}(2005){Adami}, {Biviano}, {Durret}, \&
  {Mazure}}]{Adami+05}
{Adami}, C., {Biviano}, A., {Durret}, F., \& {Mazure}, A. 2005, \aap, 443, 17

\bibitem[{{Allen} {et~al.}(2008){Allen}, {Rapetti}, {Schmidt}, {Ebeling},
  {Morris}, \& {Fabian}}]{Allen+08}
{Allen}, S.~W., {Rapetti}, D.~A., {Schmidt}, R.~W., {et~al.} 2008, \mnras, 383,
  879

\bibitem[{{Anders} \& {Grevesse}(1989)}]{AG89}
{Anders}, E. \& {Grevesse}, N. 1989, \gca, 53, 197

\bibitem[{{Balucinska-Church} \& {McCammon}(1992)}]{BCM92}
{Balucinska-Church}, M. \& {McCammon}, D. 1992, \apj, 400, 699

\bibitem[{{Bardeau} {et~al.}(2005){Bardeau}, {Kneib}, {Czoske}, {Soucail},
  {Smail}, {Ebeling}, \& {Smith}}]{Bardeau+05}
{Bardeau}, S., {Kneib}, J., {Czoske}, O., {et~al.} 2005, \aap, 434, 433

\bibitem[{{Barrena} {et~al.}(2002){Barrena}, {Biviano}, {Ramella}, {Falco}, \&
  {Seitz}}]{Barrena+02}
{Barrena}, R., {Biviano}, A., {Ramella}, M., {Falco}, E.~E., \& {Seitz}, S.
  2002, \aap, 386, 816

\bibitem[{{Bartelmann}(1996)}]{Bartelmann96}
{Bartelmann}, M. 1996, \aap, 313, 697

\bibitem[{{Beers} {et~al.}(1990){Beers}, {Flynn}, \& {Gebhardt}}]{BFG90}
{Beers}, T.~C., {Flynn}, K., \& {Gebhardt}, K. 1990, \aj, 100, 32

\bibitem[{{Bhattacharya} {et~al.}(2013){Bhattacharya}, {Habib}, {Heitmann}, \&
  {Vikhlinin}}]{BHHV13}
{Bhattacharya}, S., {Habib}, S., {Heitmann}, K., \& {Vikhlinin}, A. 2013, \apj,
  766, 32

\bibitem[{{Binney} \& {Tremaine}(1987)}]{BT87}
{Binney}, J. \& {Tremaine}, S. 1987, Galactic dynamics (Princeton, NJ,
  Princeton University Press, 1987, 747 p.)

\bibitem[{{Biviano}(2006)}]{Biviano06b}
{Biviano}, A. 2006, in EAS Publications Series, ed. G.~A. {Mamon}, F.~{Combes},
  C.~{Deffayet}, \& B.~{Fort}, 171--178

\bibitem[{{Biviano} {et~al.}(1996){Biviano}, {Durret}, {Gerbal}, {Le Fevre},
  {Lobo}, {Mazure}, \& {Slezak}}]{Biviano+96}
{Biviano}, A., {Durret}, F., {Gerbal}, D., {et~al.} 1996, \aap, 311, 95

\bibitem[{{Biviano} \& {Girardi}(2003)}]{BG03}
{Biviano}, A. \& {Girardi}, M. 2003, \apj, 585, 205

\bibitem[{{Biviano} \& {Katgert}(2004)}]{BK04}
{Biviano}, A. \& {Katgert}, P. 2004, \aap, 424, 779

\bibitem[{{Biviano} {et~al.}(2006){Biviano}, {Murante}, {Borgani}, {Diaferio},
  {Dolag}, \& {Girardi}}]{Biviano+06}
{Biviano}, A., {Murante}, G., {Borgani}, S., {et~al.} 2006, \aap, 456, 23

\bibitem[{{Biviano} {et~al.}(2013){Biviano}, {Rosati}, {Balestra}, {Mercurio},
  {Girardi}, {Nonino}, {Grillo}, {Scodeggio}, {Lemze}, {Kelson}, {Umetsu},
  {Postman}, {Zitrin}, {Czoske}, {Ettori}, {Fritz}, {Lombardi}, {Maier},
  {Medezinski}, {Mei}, {Presotto}, {Strazzullo}, {Tozzi}, {Ziegler},
  {Annunziatella}, {Bartelmann}, {Benitez}, {Bradley}, {Brescia}, {Broadhurst},
  {Coe}, {Demarco}, {Donahue}, {Ford}, {Gobat}, {Graves}, {Koekemoer},
  {Kuchner}, {Melchior}, {Meneghetti}, {Merten}, {Moustakas}, {Munari}, {Reg{\H
  o}s}, {Sartoris}, {Seitz}, \& {Zheng}}]{Biviano+13}
{Biviano}, A., {Rosati}, P., {Balestra}, I., {et~al.} 2013, \aap, 558, A1

\bibitem[{{Biviano} \& {Salucci}(2006)}]{BS06}
{Biviano}, A. \& {Salucci}, P. 2006, \aap, 452, 75

\bibitem[{{B{\"o}hringer} {et~al.}(2010){B{\"o}hringer}, {Pratt}, {Arnaud},
  {Borgani}, {Croston}, {Ponman}, {Ameglio}, {Temple}, \&
  {Dolag}}]{Bohringer+10}
{B{\"o}hringer}, H., {Pratt}, G.~W., {Arnaud}, M., {et~al.} 2010, \aap, 514,
  A32

\bibitem[{{Borgani} \& {Guzzo}(2001)}]{BG01}
{Borgani}, S. \& {Guzzo}, L. 2001, \nat, 409, 39

\bibitem[{{Briel} {et~al.}(1991){Briel}, {Henry}, {Schwarz}, {Bohringer},
  {Ebeling}, {Edge}, {Hartner}, {Schindler}, {Trumper}, \& {Voges}}]{Briel+91}
{Briel}, U.~G., {Henry}, J.~P., {Schwarz}, R.~A., {et~al.} 1991, \aap, 246, L10

\bibitem[{{Cavaliere} \& {Fusco-Femiano}(1976)}]{CFF76}
{Cavaliere}, A. \& {Fusco-Femiano}, R. 1976, \aap, 49, 137

\bibitem[{{Coe} {et~al.}(2010){Coe}, {Ben{\'{\i}}tez}, {Broadhurst}, \&
  {Moustakas}}]{Coe+10}
{Coe}, D., {Ben{\'{\i}}tez}, N., {Broadhurst}, T., \& {Moustakas}, L.~A. 2010,
  \apj, 723, 1678

\bibitem[{{Conroy} \& {Ostriker}(2008)}]{CO08}
{Conroy}, C. \& {Ostriker}, J.~P. 2008, \apj, 681, 151

\bibitem[{{Cypriano} {et~al.}(2004){Cypriano}, {Sodr{\'e}}, {Kneib}, \&
  {Campusano}}]{Cypriano+04}
{Cypriano}, E.~S., {Sodr{\'e}}, Jr., L., {Kneib}, J.-P., \& {Campusano}, L.~E.
  2004, \apj, 613, 95

\bibitem[{{De Boni} {et~al.}(2013){De Boni}, {Ettori}, {Dolag}, \&
  {Moscardini}}]{DeBoni+13}
{De Boni}, C., {Ettori}, S., {Dolag}, K., \& {Moscardini}, L. 2013, \mnras,
  428, 2921

\bibitem[{{DeCarlo}(1997)}]{decarlo97}
{DeCarlo}, L.~T. 1997, Psychological Methods, 2, 292

\bibitem[{{den Hartog} \& {Katgert}(1996)}]{dHK96}
{den Hartog}, R. \& {Katgert}, P. 1996, \mnras, 279, 349

\bibitem[{{Diaferio}(1999)}]{Diaferio99}
{Diaferio}, A. 1999, \mnras, 309, 610

\bibitem[{{Diaferio} \& {Geller}(1997)}]{DG97}
{Diaferio}, A. \& {Geller}, M.~J. 1997, \apj, 481, 633

\bibitem[{{Dressler} \& {Shectman}(1988)}]{DS88}
{Dressler}, A. \& {Shectman}, S.~A. 1988, \aj, 95, 985

\bibitem[{{Dubois} {et~al.}(2013){Dubois}, {Pichon}, {Devriendt}, {Silk},
  {Haehnelt}, {Kimm}, \& {Slyz}}]{Dubois+13}
{Dubois}, Y., {Pichon}, C., {Devriendt}, J., {et~al.} 2013, \mnras, 428, 2885

\bibitem[{{Eckert} {et~al.}(2013){Eckert}, {Ettori}, {Molendi}, {Vazza}, \&
  {Paltani}}]{Eckert+13}
{Eckert}, D., {Ettori}, S., {Molendi}, S., {Vazza}, F., \& {Paltani}, S. 2013,
  \aap, 551, A23

\bibitem[{{El{\'{\i}}asd{\'o}ttir} {et~al.}(2007){El{\'{\i}}asd{\'o}ttir},
  {Limousin}, {Richard}, {Hjorth}, {Kneib}, {Natarajan}, {Pedersen}, {Jullo},
  \& {Paraficz}}]{Eliasdottir+07}
{El{\'{\i}}asd{\'o}ttir}, {\'A}., {Limousin}, M., {Richard}, J., {et~al.} 2007,
  arXiv:astro-ph/0710.5636, Vol. 710

\bibitem[{{Escalera} {et~al.}(1994){Escalera}, {Biviano}, {Girardi},
  {Giuricin}, {Mardirossian}, {Mazure}, \& {Mezzetti}}]{Escalera+94}
{Escalera}, E., {Biviano}, A., {Girardi}, M., {et~al.} 1994, \apj, 423, 539

\bibitem[{{Frederiksen} {et~al.}(2009){Frederiksen}, {Hansen}, {Host}, \&
  {Roncadelli}}]{FHHR09}
{Frederiksen}, T.~F., {Hansen}, S.~H., {Host}, O., \& {Roncadelli}, M. 2009,
  \apj, 700, 1603

\bibitem[{{Fukugita} {et~al.}(1998){Fukugita}, {Hogan}, \& {Peebles}}]{FHP98}
{Fukugita}, M., {Hogan}, C.~J., \& {Peebles}, P. J.~E. 1998, \apj, 503, 518

\bibitem[{{Geller} {et~al.}(2013){Geller}, {Diaferio}, {Rines}, \&
  {Serra}}]{GDRS13}
{Geller}, M.~J., {Diaferio}, A., {Rines}, K.~J., \& {Serra}, A.~L. 2013, \apj,
  764, 58

\bibitem[{{Gifford} {et~al.}(2013){Gifford}, {Miller}, \& {Kern}}]{Gifford+13}
{Gifford}, D., {Miller}, C., \& {Kern}, N. 2013, \apj, 773, 116

\bibitem[{{Girardi} \& {Biviano}(2002)}]{GB02}
{Girardi}, M. \& {Biviano}, A. 2002, Optical Analysis of Cluster Mergers (ASSL
  Vol.~272: Merging Processes in Galaxy Clusters), 39--77

\bibitem[{{Guennou} {et~al.}(2013){Guennou}, {Adami}, {Durret}, {Lima Neto},
  {Ulmer}, {Clowe}, {LeBrun}, {Martinet}, {Allam}, {Annis}, {Basa}, {Benoist},
  {Biviano}, {Cappi}, {Cypriano}, {Gavazzi}, {Halliday}, {Ilbert}, {Jullo},
  {Just}, {Limousin}, {M{\'a}rquez}, {Mazure}, {Murphy}, {Plana}, {Rostagni},
  {Russeil}, {Schirmer}, {Slezak}, {Tucker}, {Zaritsky}, \&
  {Ziegler}}]{Guennou+13}
{Guennou}, L., {Adami}, C., {Durret}, F., {et~al.} 2013, arXiv:1311.6922

\bibitem[{{Guennou} {et~al.}(2010){Guennou}, {Adami}, {Ulmer}, {Lebrun},
  {Durret}, {Johnston}, {Ilbert}, {Clowe}, {Gavazzi}, {Murphy}, {Schrabback},
  {Allam}, {Annis}, {Basa}, {Benoist}, {Biviano}, {Cappi}, {Kubo}, {Marshall},
  {Mazure}, {Rostagni}, {Russeil}, \& {Slezak}}]{Guennou+10}
{Guennou}, L., {Adami}, C., {Ulmer}, M.~P., {et~al.} 2010, \aap, 523, A21

\bibitem[{{Halliday} {et~al.}(2004){Halliday}, {Milvang-Jensen}, {Poirier},
  {Poggianti}, {Jablonka}, {Arag{\'o}n-Salamanca}, {Saglia}, {De Lucia},
  {Pell{\'o}}, {Simard}, {Clowe}, {Rudnick}, {Dalcanton}, {White}, \&
  {Zaritsky}}]{Halliday+04}
{Halliday}, C., {Milvang-Jensen}, B., {Poirier}, S., {et~al.} 2004, \aap, 427,
  397

\bibitem[{{Hinshaw} {et~al.}(2012){Hinshaw}, {Larson}, {Komatsu}, {Spergel},
  {Bennett}, {Dunkley}, {Nolta}, {Halpern}, {Hill}, {Odegard}, {Page}, {Smith},
  {Weiland}, {Gold}, {Jarosik}, {Kogut}, {Limon}, {Meyer}, {Tucker}, {Wollack},
  \& {Wright}}]{Hinshaw+12}
{Hinshaw}, G., {Larson}, D., {Komatsu}, E., {et~al.} 2012, arXiv:1212.5226

\bibitem[{{Jee} {et~al.}(2005){Jee}, {White}, {Ben{\'{\i}}tez}, {Ford},
  {Blakeslee}, {Rosati}, {Demarco}, \& {Illingworth}}]{Jee+05}
{Jee}, M.~J., {White}, R.~L., {Ben{\'{\i}}tez}, N., {et~al.} 2005, \apj, 618,
  46

\bibitem[{{Johnson} {et~al.}(2006){Johnson}, {Best}, {Zaritsky}, {Clowe},
  {Arag{\'o}n-Salamanca}, {Halliday}, {Jablonka}, {Milvang-Jensen},
  {Pell{\'o}}, {Poggianti}, {Rudnick}, {Saglia}, {Simard}, \&
  {White}}]{Johnson+06}
{Johnson}, O., {Best}, P., {Zaritsky}, D., {et~al.} 2006, \mnras, 371, 1777

\bibitem[{{Jullo} {et~al.}(2007){Jullo}, {Kneib}, {Limousin},
  {El{\'{\i}}asd{\'o}ttir}, {Marshall}, \& {Verdugo}}]{Jullo+07}
{Jullo}, E., {Kneib}, J.-P., {Limousin}, M., {et~al.} 2007, New Journal of
  Physics, 9, 447

\bibitem[{{Kaastra} \& {Mewe}(1993)}]{KM93}
{Kaastra}, J.~S. \& {Mewe}, R. 1993, A\&AS, 97, 443

\bibitem[{{Kalberla} {et~al.}(2005){Kalberla}, {Burton}, {Hartmann}, {Arnal},
  {Bajaja}, {Morras}, \& {P{\"o}ppel}}]{Kalberla+05}
{Kalberla}, P.~M.~W., {Burton}, W.~B., {Hartmann}, D., {et~al.} 2005, \aap,
  440, 775

\bibitem[{{Kasun} \& {Evrard}(2005)}]{KE05}
{Kasun}, S.~F. \& {Evrard}, A.~E. 2005, \apj, 629, 781

\bibitem[{{Katgert} {et~al.}(2004){Katgert}, {Biviano}, \& {Mazure}}]{KBM04}
{Katgert}, P., {Biviano}, A., \& {Mazure}, A. 2004, \apj, 600, 657

\bibitem[{{Kent} \& {Gunn}(1982)}]{KG82}
{Kent}, S.~M. \& {Gunn}, J.~E. 1982, \aj, 87, 945

\bibitem[{{Kim} {et~al.}(2002){Kim}, {Annis}, {Strauss}, \& {Lupton}}]{KASL02}
{Kim}, R.~S.~J., {Annis}, J., {Strauss}, M.~A., \& {Lupton}, R.~H. 2002, in
  Astronomical Society of the Pacific Conference Series, Vol. 268, Tracing
  Cosmic Evolution with Galaxy Clusters, ed. S.~{Borgani}, M.~{Mezzetti}, \&
  R.~{Valdarnini}, 395

\bibitem[{{Leonard} {et~al.}(2011){Leonard}, {King}, \& {Goldberg}}]{LKG11}
{Leonard}, A., {King}, L.~J., \& {Goldberg}, D.~M. 2011, \mnras, 413, 789

\bibitem[{{Liedahl} {et~al.}(1995){Liedahl}, {Osterheld}, \&
  {Goldstein}}]{LOG95}
{Liedahl}, D.~A., {Osterheld}, A.~L., \& {Goldstein}, W.~H. 1995, \apjl, 438,
  L115

\bibitem[{{Limousin} {et~al.}(2007{\natexlab{a}}){Limousin}, {Kneib},
  {Bardeau}, {Natarajan}, {Czoske}, {Smail}, {Ebeling}, \&
  {Smith}}]{Limousin+07}
{Limousin}, M., {Kneib}, J.~P., {Bardeau}, S., {et~al.} 2007{\natexlab{a}},
  \aap, 461, 881

\bibitem[{{Limousin} {et~al.}(2005){Limousin}, {Kneib}, \&
  {Natarajan}}]{Limousin+05}
{Limousin}, M., {Kneib}, J.-P., \& {Natarajan}, P. 2005, \mnras, 356, 309

\bibitem[{{Limousin} {et~al.}(2007{\natexlab{b}}){Limousin}, {Richard},
  {Jullo}, {Kneib}, {Fort}, {Soucail}, {El{\'{\i}}asd{\'o}ttir}, {Natarajan},
  {Ellis}, {Smail}, {Czoske}, {Smith}, {Hudelot}, {Bardeau}, {Ebeling},
  {Egami}, \& {Knudsen}}]{Limousin+07b}
{Limousin}, M., {Richard}, J., {Jullo}, E., {et~al.} 2007{\natexlab{b}}, \apj,
  668, 643

\bibitem[{{{\L}okas} \& {Mamon}(2001)}]{LM01}
{{\L}okas}, E.~L. \& {Mamon}, G.~A. 2001, \mnras, 321, 155

\bibitem[{{{\L}okas} \& {Mamon}(2003)}]{LM03}
{{\L}okas}, E.~L. \& {Mamon}, G.~A. 2003, \mnras, 343, 401

\bibitem[{{{\L}okas} {et~al.}(2006){{\L}okas}, {Wojtak}, {Gottl{\"o}ber},
  {Mamon}, \& {Prada}}]{Lokas+06}
{{\L}okas}, E.~L., {Wojtak}, R., {Gottl{\"o}ber}, S., {Mamon}, G.~A., \&
  {Prada}, F. 2006, \mnras, 367, 1463

\bibitem[{{Mamon} {et~al.}(2013){Mamon}, {Biviano}, \& {Bou{\'e}}}]{MBB13}
{Mamon}, G.~A., {Biviano}, A., \& {Bou{\'e}}, G. 2013, \mnras, 429, 3079

\bibitem[{{Mamon} {et~al.}(2010){Mamon}, {Biviano}, \& {Murante}}]{MBM10}
{Mamon}, G.~A., {Biviano}, A., \& {Murante}, G. 2010, \aap, 520, A30

\bibitem[{{Mamon} \& {{\L}okas}(2005)}]{ML05b}
{Mamon}, G.~A. \& {{\L}okas}, E.~L. 2005, \mnras, 363, 705

\bibitem[{{Markevitch} {et~al.}(2004){Markevitch}, {Gonzalez}, {Clowe},
  {Vikhlinin}, {Forman}, {Jones}, {Murray}, \& {Tucker}}]{Markevitch+04}
{Markevitch}, M., {Gonzalez}, A.~H., {Clowe}, D., {et~al.} 2004, \apj, 606, 819

\bibitem[{{Markevitch} {et~al.}(2002){Markevitch}, {Gonzalez}, {David},
  {Vikhlinin}, {Murray}, {Forman}, {Jones}, \& {Tucker}}]{Markevitch+02}
{Markevitch}, M., {Gonzalez}, A.~H., {David}, L., {et~al.} 2002, \apjl, 567,
  L27

\bibitem[{{Mohr} {et~al.}(1993){Mohr}, {Fabricant}, \& {Geller}}]{MFG93}
{Mohr}, J.~J., {Fabricant}, D.~G., \& {Geller}, M.~J. 1993, \apj, 413, 492

\bibitem[{{Navarro} {et~al.}(1997){Navarro}, {Frenk}, \& {White}}]{NFW97}
{Navarro}, J.~F., {Frenk}, C.~S., \& {White}, S. D.~M. 1997, \apj, 490, 493

\bibitem[{{Nelson} {et~al.}(2001){Nelson}, {Gonzalez}, {Zaritsky}, \&
  {Dalcanton}}]{Nelson+01}
{Nelson}, A.~E., {Gonzalez}, A.~H., {Zaritsky}, D., \& {Dalcanton}, J.~J. 2001,
  \apj, 563, 629

\bibitem[{{Neumann} {et~al.}(2001){Neumann}, {Arnaud}, {Gastaud}, {Aghanim},
  {Lumb}, {Briel}, {Vestrand}, {Stewart}, {Molendi}, \& {Mittaz}}]{Neumann+01}
{Neumann}, D.~M., {Arnaud}, M., {Gastaud}, R., {et~al.} 2001, \aap, 365, L74

\bibitem[{{O'Hara} {et~al.}(2004){O'Hara}, {Mohr}, \& {Guerrero}}]{OHara+04}
{O'Hara}, T.~B., {Mohr}, J.~J., \& {Guerrero}, M.~A. 2004, \apj, 604, 604

\bibitem[{{Planck Collaboration} {et~al.}(2013{\natexlab{a}}){Planck
  Collaboration}, {Ade}, {Aghanim}, {Armitage-Caplan}, {Arnaud}, {Ashdown},
  {Atrio-Barandela}, {Aumont}, {Baccigalupi}, {Banday}, \&
  et~al.}]{PlanckCollaboration+13}
{Planck Collaboration}, {Ade}, P.~A.~R., {Aghanim}, N., {et~al.}
  2013{\natexlab{a}}, arXiv:1303.5076

\bibitem[{{Planck Collaboration} {et~al.}(2013{\natexlab{b}}){Planck
  Collaboration}, {Ade}, {Aghanim}, {Armitage-Caplan}, {Arnaud}, {Ashdown},
  {Atrio-Barandela}, {Aumont}, {Baccigalupi}, {Banday}, \& et~al.}]{Planck13}
{Planck Collaboration}, {Ade}, P.~A.~R., {Aghanim}, N., {et~al.}
  2013{\natexlab{b}}, arXiv:1303.5080

\bibitem[{{Pratt} {et~al.}(2005){Pratt}, {B{\"o}hringer}, \&
  {Finoguenov}}]{PBF05}
{Pratt}, G.~W., {B{\"o}hringer}, H., \& {Finoguenov}, A. 2005, \aap, 433, 777

\bibitem[{{Ramella} {et~al.}(2007){Ramella}, {Biviano}, {Pisani}, {Varela},
  {Bettoni}, {Couch}, {D'Onofrio}, {Dressler}, {Fasano}, {Kj{\o}rgaard},
  {Moles}, {Pignatelli}, \& {Poggianti}}]{Ramella+07}
{Ramella}, M., {Biviano}, A., {Pisani}, A., {et~al.} 2007, \aap, 470, 39

\bibitem[{{Rasia} {et~al.}(2006){Rasia}, {Ettori}, {Moscardini}, {Mazzotta},
  {Borgani}, {Dolag}, {Tormen}, {Cheng}, \& {Diaferio}}]{Rasia+06}
{Rasia}, E., {Ettori}, S., {Moscardini}, L., {et~al.} 2006, \mnras, 369, 2013

\bibitem[{{Rhee} \& {Katgert}(1987)}]{RK87}
{Rhee}, G.~F.~R.~N. \& {Katgert}, P. 1987, \aap, 183, 217

\bibitem[{{Sanchis} {et~al.}(2004){Sanchis}, {{\L}okas}, \& {Mamon}}]{SLM04}
{Sanchis}, T., {{\L}okas}, E.~L., \& {Mamon}, G.~A. 2004, \mnras, 347, 1198

\bibitem[{{Serna} \& {Gerbal}(1996)}]{SG96}
{Serna}, A. \& {Gerbal}, D. 1996, \aap, 309, 65

\bibitem[{{Serra} {et~al.}(2011){Serra}, {Diaferio}, {Murante}, \&
  {Borgani}}]{Serra+11}
{Serra}, A.~L., {Diaferio}, A., {Murante}, G., \& {Borgani}, S. 2011, \mnras,
  412, 800

\bibitem[{{Silverman}(1986)}]{Silverman86}
{Silverman}, B.~W. 1986, Density estimation for statistics and data analysis

\bibitem[{{Tiret} {et~al.}(2007){Tiret}, {Combes}, {Angus}, {Famaey}, \&
  {Zhao}}]{Tiret+07}
{Tiret}, O., {Combes}, F., {Angus}, G.~W., {Famaey}, B., \& {Zhao}, H.~S. 2007,
  \aap, 476, L1

\bibitem[{{van der Marel} {et~al.}(2000){van der Marel}, {Magorrian},
  {Carlberg}, {Yee}, \& {Ellingson}}]{vanderMarel+00}
{van der Marel}, R.~P., {Magorrian}, J., {Carlberg}, R.~G., {Yee}, H.~K.~C., \&
  {Ellingson}, E. 2000, \aj, 119, 2038

\bibitem[{{Wojtak}(2013)}]{Wojtak13}
{Wojtak}, R. 2013, arXiv:1310.3624

\bibitem[{{Wojtak} \& {{\L}okas}(2010)}]{WL10}
{Wojtak}, R. \& {{\L}okas}, E.~L. 2010, \mnras, 408, 2442

\bibitem[{{Wojtak} {et~al.}(2009){Wojtak}, {{\L}okas}, {Mamon}, \&
  {Gottl{\"o}ber}}]{Wojtak+09}
{Wojtak}, R., {{\L}okas}, E.~L., {Mamon}, G.~A., \& {Gottl{\"o}ber}, S. 2009,
  \mnras, 399, 812

\bibitem[{{Wojtak} {et~al.}(2007){Wojtak}, {{\L}okas}, {Mamon},
  {Gottl{\"o}ber}, {Prada}, \& {Moles}}]{Wojtak+07}
{Wojtak}, R., {{\L}okas}, E.~L., {Mamon}, G.~A., {et~al.} 2007, \aap, 466, 437

\bibitem[{{Zhang} {et~al.}(2010){Zhang}, {Okabe}, {Finoguenov}, {Smith},
  {Piffaretti}, {Valdarnini}, {Babul}, {Evrard}, {Mazzotta}, {Sanderson}, \&
  {Marrone}}]{Zhang+10}
{Zhang}, Y.-Y., {Okabe}, N., {Finoguenov}, A., {et~al.} 2010, \apj, 711, 1033

\end{thebibliography}

\end{document}